\documentclass[aps,pre,twocolumn,preprintnumbers,amsmath,amssymb,floatfix,superscriptaddress,nofootinbib]{revtex4-1}
\usepackage{graphicx}
\usepackage{bm}
\usepackage{color}
\usepackage[normalem]{ulem} % sout

\def\nablab{{\bm \nabla}}

\def\J{\mathcal{J}}
\def\O{\mathcal{O}}
\def\enr{\mathcal{E}}

\def\eq{{\rm ref}}

\def\figures{.}

\definecolor{gray}{rgb}{0.5,0.5,0.5}
\definecolor{lgray}{rgb}{0.8,0.8,0.8}
\definecolor{dgray}{rgb}{0.6,0.6,0.6}
\definecolor{dred}{rgb}{0.5,0.0,0.0}
\definecolor{dgreen}{rgb}{0.0,0.5,0.0}
\definecolor{dblue}{rgb}{0.0,0.0,0.5}
\definecolor{violet}{rgb}{0.7,0.0,0.5}

\definecolor{lred}{rgb}{1.0,0.5,0.5}
\definecolor{lgreen}{rgb}{0.5,1.0,0.5}
\definecolor{lblue}{rgb}{0.5,0.5,1.0}

\newcommand{\cg}{\color{dgreen} }
\newcommand{\cred}{\color{red} }
\newcommand{\cgray}{\color{dgray} }

\def\L{{\mathcal L}}
\def\H{{\mathcal H}}
\def\O{{\mathcal O}}

\def\J{{\mathcal J}}
\def\P{{\mathcal P}}
\def\E{{\mathcal E}}

\makeatother
\begin{document}

%\preprint{}

\title{Testing the conservative character of particle simulations: \\ I. Canonical and noncanonical guiding center model in Boozer coordinates}

\author{A.~Bierwage}
\email[]{bierwage.andreas@qst.go.jp}
\affiliation{QST, Rokkasho Fusion Institute, Aomori 039-3212, Japan}
\affiliation{QST, Naka Fusion Institute, Ibaraki 311-0193, Japan}
\author{R.B.~White}
\affiliation{PPPL, Princeton University, Princeton NJ 08543, USA}
\author{A.~Matsuyama}
\affiliation{QST, Rokkasho Fusion Institute, Aomori 039-3212, Japan}

\date{\today}

\begin{abstract} % max. 600 characters (currently 588 with spaces, 499 without)
The guiding center (GC) Lagrangian in Boozer coordinates for toroidally confined plasmas can be cast into canonical form by eliminating a term containing the covariant component $B_{\Psi_{\rm P}}$ of the magnetic field vector with respect to the poloidal flux function $\Psi_{\rm P}$. Considering fast ions in the presence of a shear Alfv\'{e}n wave field with fixed amplitude, fixed frequency and a single toroidal mode number $n$, we show that simulations using the code {\tt ORBIT} with and without $B_{\Psi_{\rm P}}$ yield practically the same resonant and nonresonant GC orbits. The numerical results are consistent with theoretical analyses (presented in the Appendix), which show that the unabridged GC Lagrangian with $B_{\Psi_{\rm P}}$ retained yields equations of motion that possess two key properties of Hamiltonian flows: (i) phase space conservation, and (ii) energy conservation. As counter-examples, we also show cases where energy conservation (ii) or both conservation laws (i) \& (ii) are broken by omitting certain small terms. When testing the conservative character of the simulation code, it is found to be beneficial to apply perturbations that do not resemble normal (eigen)modes of the plasma. The deviations are enhanced and, thus, more easily spotted when one inspects wave-particle interactions using nonnormal modes.
\end{abstract}

\maketitle

\thispagestyle{empty}
\everypar{\looseness=-1} % squeeze space

% =============================================================================
\section{Introduction}
\label{sec:intro}

The guiding center (GC) model of charged particle motion in a magnetized plasma enjoys great popularity in the study of confinement and transport phenomena in laboratory and space plasmas. In the realm of tokamak and stellarator plasmas, it is often advantageous to work with equations of motion in {\it unperturbed} magnetic coordinates. Although these coordinates are usually noncanonical (their Jacobians are not constants), it was found that the GC phase-space Lagrangian expressed in Boozer coordinates \cite{Boozer81, Boozer82} can be cast into canonical form \cite{White84}, which guarantees that the GC model inherits the Hamiltonian nature of the overlying Newton-Lorentz equation for charged particle motion in a fluctuating electromagnetic field that satisfies Maxwell's equations. This is important for an accurate analysis of resonances and their role for magnetic confinement \cite{White18,White21b,White22}. The formulation we consider here uses prescribed fields, whose fluctuations are not affected by the charged particles being simulated. Moreover, the fields are given in terms of the scalar and vector potentials $\Phi$ and ${\bm A}$, so that the model is guaranteed to satisfy Maxwell's equations via the built-in definitions ${\bm E} = -\nablab\Phi - \partial_t{\bm A}$ and ${\bm B} = \nablab\times{\bm A}$. (Models where the fluctuations are expressed in terms of the physical fields ${\bm E}$ and ${\bm B}$ are discussed in a companion paper \cite{Bierwage22c}.)

This Hamiltonian GC model has been adopted in several simulation codes, starting with {\tt ORBIT} \cite{White84}. The modeling step that puts the symplectic part of the Lagrangian into canonical form is essentially a truncation, assuming that certain coefficients are small. In this work, we study the role of the truncated terms and present numerical and analytical evidence showing that it is safe to retain them. One reason for doing so is that the smallness of those terms is not always guaranteed.

We begin with a short review of the formalism, details of which can be found in Appendix~\ref{apdx:theory}. The magnetic field strength $B$ is normalized by a convenient reference value $B_0$ and time $t$ is normalized by the corresponding gyrofrequency $\Omega_{\rm g0} = ZeB_0/M$ for particles with electric charge $Ze$ and mass $M$ (in SI units). Energy is normalized by $M\Omega_{\rm g0}^2 = ZeB_0\Omega_{\rm g0}$. The GC phase-space Lagrangian is then
\begin{equation}
\L({\bm X}_{\rm gc},\rho_\parallel,\mu,\theta,\dot{\bm X}_{\rm gc},\dot{\theta},t) = ({\bm A} + \rho_\parallel{\bm B})\cdot\dot{\bm X}_{\rm gc} + \mu\dot{\theta} - \H,
\label{eq:l}
\end{equation}

\noindent where ${\bm B} = \nablab\times{\bm A}$ is the magnetic field vector, $\dot{\bm X}_{\rm gc} \equiv {\rm d}{\bm X}_{\rm gc}/{\rm d}t$ is the GC velocity vector, and $\rho_\parallel \equiv {\bm X}_{\rm gc}\cdot{\bm B}/(B\Omega_{\rm g}) = u/\Omega_{\rm g}$ is its parallel component divided by the gyrofrequency, which is $\Omega_{\rm g} = B$ in normalized units. The Hamiltonian $\H({\bm X}_{\rm gc},\rho_\parallel,\mu,t)$ can be written as \cite{Cary09}
\begin{equation}
\H = \mu B + \Phi^*, \quad \Phi^* = \Phi + \rho_\parallel^2 B^2/2 {\cgray - |{\bm v}_{\rm E}^2|/2},
\label{eq:h}
\end{equation}

\noindent where we retained for completeness the term $-|{\bm v}_{\rm E}|^2/2 = -|{\bm E}_\perp|^2/(2B^2)$, where ${\bm v}_{\rm E} = {\bm E}\times{\bm B}/B^2$ is the electric drift velocity associated with the fluctuating electric field ${\bm E} = -\nablab\Phi - \partial_t\delta{\bm A}$. This term is reminiscent of a not-yet-averaged ponderomotive potential (see the Appendix for further discussion). It is printed gray in Eq.~(\ref{eq:h}) since it is not currently included in {\tt ORBIT}.

All variables are taken to be independent of the gyrophase $\theta$ (which satisfies $\dot{\theta} = \Omega_{\rm g}$), so that the magnetic moment $\mu$ in this model is conserved exactly. The magnetic field vector ${\bm B}(t) = {\bm B}_\eq + \delta{\bm B}(t) = \nablab\times({\bm A}_\eq + \delta{\bm A})$ contains a time-independent reference field ${\bm B}_\eq$, whose contravariant form in generic straight-field-line coordinates $(\Psi,\vartheta,\zeta)$ is
\begin{equation}
{\bm B}_\eq = \nablab\times(\Psi\nablab\vartheta - \Psi_{\rm P}\nablab\zeta).
\label{eq:b_flux}
\end{equation}

\noindent where $\Psi$ and $\Psi_{\rm P}$ are the poloidal and toroidal fluxes,\footnote{Strictly speaking, our $\Psi$ and $\Psi_{\rm P}$ are the fluxes divided by $2\pi$.}
and $\vartheta$ and $\zeta$ are the poloidal and toroidal angle coordinates. The magnetic field line helicity is measured by the tokamak safety factor $q(\Psi_{\rm P}) = {\rm d}\Psi/{\rm d}\Psi_{\rm P}$. Assuming axisymmetry ($\partial_\zeta\Psi = 0$), the reference field has the following covariant representation in Boozer coordinates:
\begin{equation}
{\bm B}_\eq = g(\Psi_{\rm P})\nablab\zeta + I(\Psi_{\rm P})\nablab\vartheta + q\beta_*(\Psi_{\rm P},\vartheta)\nablab\Psi_{\rm P}.
\label{eq:b_boozer}
\end{equation}

\noindent We constrain the fluctuating component $\delta{\bm B} = \nablab\times\delta{\bm A}$ of the magnetic field to perturbations of the form $\delta{\bm A} = \alpha{\bm B}_\eq$. The scalar function $\alpha(\Psi_{\rm P},\vartheta,\zeta,t) = \delta{A}_\parallel/B_\eq$ is assumed to be small, obeying $\alpha/L_B \ll 1$, where $L_B$ is the scale length of the reference field's nonuniformities (gradient, curvature). Omitting terms of order $\O(\rho_\parallel^2 \alpha/L_B)$ and higher (Appendix~\ref{apdx:theory_gce}), the Lagrangian becomes \cite{White84}
\begin{equation}
\L = (\rho_\parallel + \alpha)q \beta_*\dot{\Psi}_{\rm P} + \P_\zeta \dot{\zeta} + \P_\vartheta\dot{\vartheta} + \mu\dot{\theta} - \H,
\label{eq:l_booz}
\end{equation}

\noindent with $\P_\zeta = (\rho_\parallel + \alpha)g + \Psi_{\rm P}$ and $\P_\vartheta = (\rho_\parallel + \alpha)I + \Psi$.

The phase space is 6-dimensional, so the symplectic part of the Lagrangian has canonical form when it contains the time derivatives of only three variables. It has been customary to retain the products of the three action variables $(\P_\zeta,\P_\vartheta,\mu)$ with the time derivatives of the three angles $(\zeta,\vartheta,\theta)$ appearing in Eq.~(\ref{eq:l_booz}) and eliminate the term $(\rho_\parallel + \alpha)\beta_*\dot{\Psi}$. This can be realized in several ways \cite{White84, White90, White03, WhiteTokBook3, Cary09}, all of which rely on $\beta_* = {\bm B}_\eq\cdot\partial{\bm x}/\partial\Psi = -I\nablab\Psi\cdot\nablab\vartheta/|\nablab\Psi|^2$ being small. All terms containing $\beta_*$ can then be consistently omitted along with other higher-order terms in the small parameter $\epsilon = {\rm max}\{\omega/\Omega_{\rm g},\rho_{\rm g}/L_B\}$, where $\omega$ is the angular frequency of the field perturbation at hand, and $\rho_{\rm g} = v_\perp/\Omega_{\rm g}$ is the gyroradius.

It should be noted that even the system excluding $\beta_*$ is truly Hamiltonian only if the electromagnetic fields are prescribed and, thus, not affected by the presence of the charged particles nor by their motion. This condition is satisfied in the setup we consider here.\footnote{Otherwise, the electromagnetic fields would also have to be treated as Hamiltonian variables in order to ensure overall energy conservation. Reduced models of wave-particle interactions can also be constructed to conserve energy \protect\cite{Pinches98} and are often used for studies of nonlinear frequency chirping (e.g., see Refs.~\protect\onlinecite{White19, Bierwage21}).}

\begin{figure}[tbp]
\centering%\vspace{-0.05cm}
\includegraphics[width=0.48\textwidth]{\figures/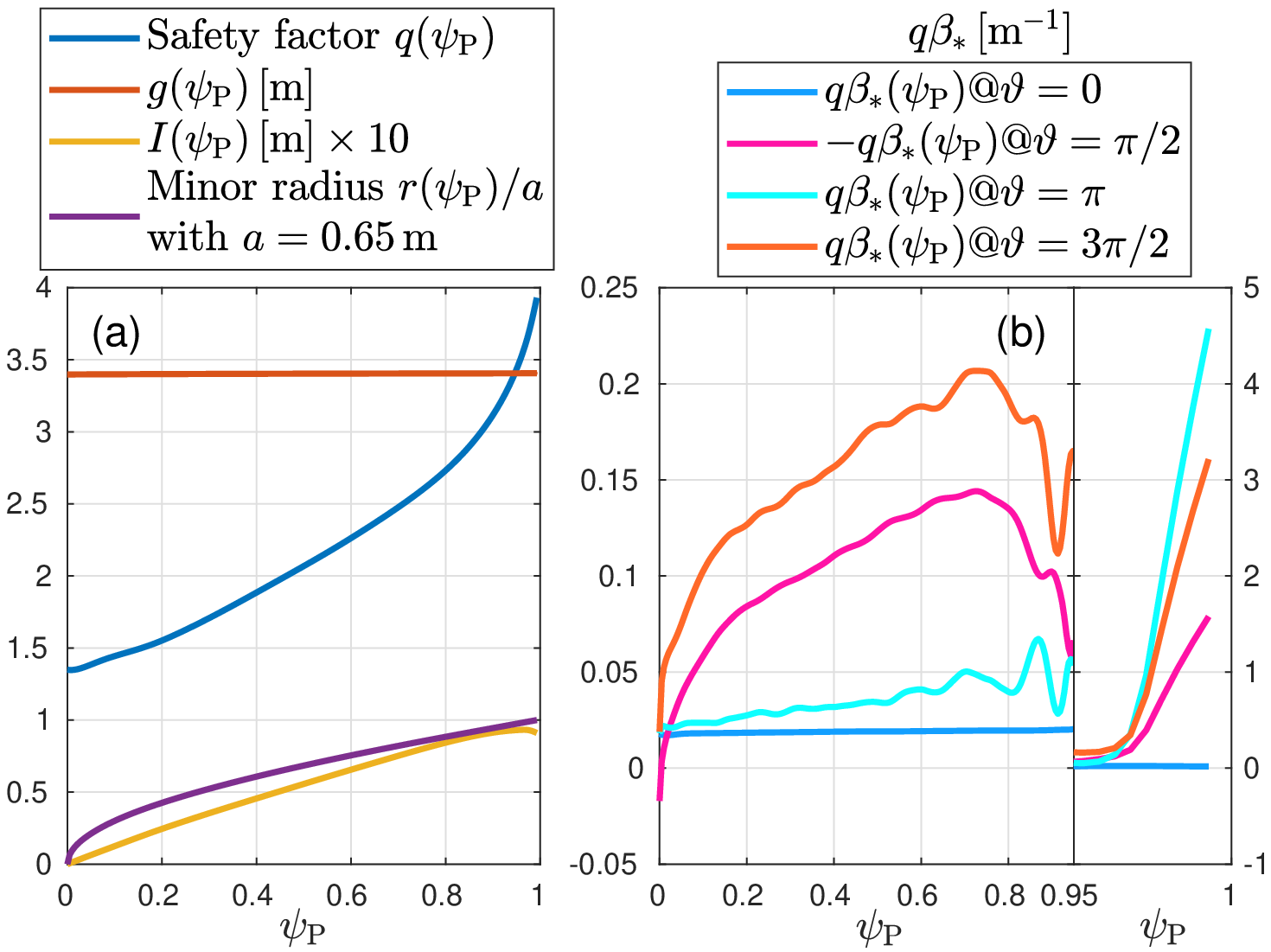}
\includegraphics[width=0.48\textwidth]{\figures/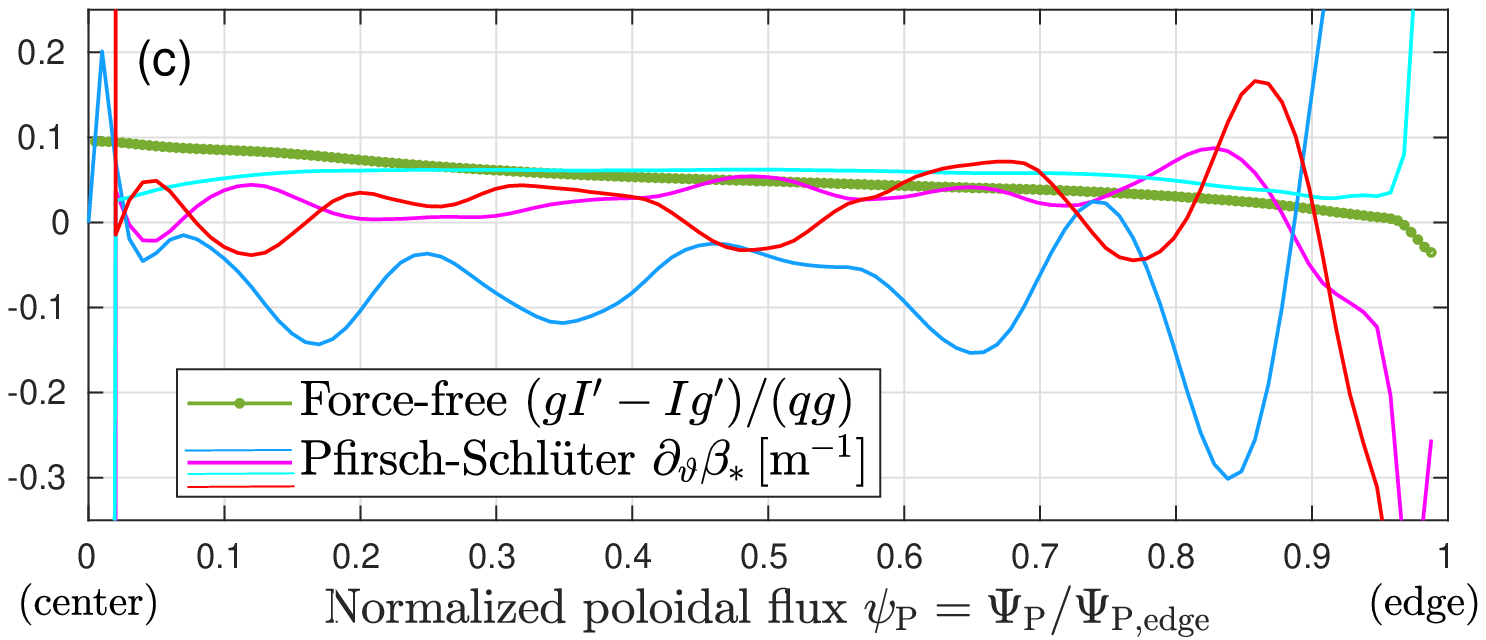}%\vspace{-0.05cm}
\caption{Equilibrium profiles for the JT-60U scenario studied in Ref.~\protect\onlinecite{Bierwage18}, which also serves us as a working example in this study. Panel (b) indicates that $q\beta_*$ may be inaccurate near the magnetic axis ($\psi_{\rm P} \rightarrow 0$), which constitutes a singularity of the polar coordinates. Low accuracy is also to be expected near the last closed flux surface ($\psi_{\rm P} \rightarrow 1$), where $q$ diverges and the resolution of our mesh appears to become insufficient. We will focus on GC orbits in the region $0.1\lesssim \psi_{\rm P} \lesssim 0.95$. Here $g$, $I$ and $\beta_*$ are normalized by $B_0 = 1.16\,{\rm T}$.}%\vspace{-0.05cm}
\label{fig:01}%
\end{figure}

To our knowledge, the effect of retaining $\beta_*$ in GC orbit-following codes has not been studied in detail. One reason for our curiosity is that the neglect of small terms may undermine the original goal to construct a model of GC motion that faithfully describes their motion on long time scales. In the case of energetic ions in a tokamak plasma, the dominant terms are usually done with their work in a matter of microseconds. If the simulation is run up to the millisecond scale (beyond which collisions become important), one should consider to include small terms, especially if they could lead to secular motion. However, it can be readily verified (Appendix~\ref{apdx:theory_erot}) that the retention of $\beta_*$ does not cause secular motion in the present GC theory, so it may be safely omitted if small.

Our second motivation is that the smallness of $\beta_*$ is disputable. Note that $\beta_*$ is multiplied by the safety factor $q$ when ${\bm B}_\eq$ and $\L$ are expressed in terms of $\Psi_{\rm P}$, as we have done in Eqs.~(\ref{eq:b_boozer}) and (\ref{eq:l_booz}). As a working example we consider the JT-60U tokamak plasma studied in Ref.~\onlinecite{Bierwage18}, for which Fig.~\ref{fig:01} shows the profiles of the relevant geometric coefficients. Figure~\ref{fig:01}(b) shows that we are dealing with values $q\beta_* \lesssim 0.2\,{\rm m^{-1}}$. This is not far from unity, so it seems reasonable to check whether the inclusion of $\beta_*$ makes a quantitative difference.

Figure~\ref{fig:01}(c) shows that the contribution of $\beta_*$ to the parallel current density $\mu_0 J_\parallel/B_\eq \approx (gI' - Ig')/(qg) - \partial_\vartheta\beta_*$ is comparable to or larger than the magnitude of the force-free part $(gI' - Ig')/(qg)$. In the region $\psi_{\rm P} > 0.95$ of the normalized poloidal flux $\psi_{\rm P} \equiv \Psi_{\rm P}/\Psi_{\rm P,edge}$, the value of $\partial_\vartheta\beta_*$ exceeds unity. As noted in Ref.~\onlinecite{Boozer81}, $\beta_*$ is related to the Pfirsch-Schl\"{u}ter current, which often leads to unphysical artifacts within the ideal MHD model (see Section 2.4 in Ref.~\onlinecite{WhiteTokBook3}). These artifacts can be eliminated by omitting $\beta_*$, which is compulsory for stellarators. In some sense, the handling of $\beta_*$ is not so much a physics problem, but an issue of mathematical consistency between the solution of the Grad-Shafranov equation (here ${\bm B}_\eq$) and the GC model when they are expressed in Boozer coordinates. In an axisymmetric diverted equilibrium like that in Fig.~\ref{fig:01}, the only problematic regions are the magnetic axis \cite{KuoPetravic83} and the separatrix, which we shall avoid here. Moreover, we constrain the parallel velocity as $|u| \ll \Omega_{\rm g}B_\eq/(\mu_0 J_\parallel) \sim q R_0 \Omega_{\rm g}$ to avoid problematic behavior \cite{CorreaRestrepo85,Burby17} of the GC Jacobian $B^*_\parallel \approx B_\eq + (\rho_\parallel+\alpha)\mu_0 J_\parallel/B_\eq$ in Eq.~(\ref{eq:bstar_approx}) \cite{Littlejohn83}.

In the present work, we consider the motion of GCs in the presence of a wave field with fixed amplitude, fixed frequency $\omega$, and single toroidal mode number $n$, so that $\Phi \propto \exp(in\zeta-i\omega t) + {\rm c.c.}$ Hamilton's equations of motion then imply that $\dot{\E}' = 0$, where $\E' = \E - \omega \P_\zeta/n$ and $\E = \H$ \cite{Hsu94}. In other words, Hamiltonian GC theory constrains the motion of GCs to invariant toroidal surfaces with $\mu = {\rm const}$.\ and $\E' = {\rm const}$., where $\E'$ is the particle energy in the frame of reference moving with the prescribed electromagnetic wave. A slightly generalized formulation of this conservation law, encompassing also the term $-|{\bm v}_{\rm E}|^2$, is given in Appendix~\ref{apdx:theory_erot}.

As a working example, we consider the interaction of fast deuterons with shear Alfv\'{e}n modes in realistic tokamak geometry but with arbitrary mode structures, including `nonnormal modes' that may arise in the course of nonlinear dynamics or due to external forcing. The mode frequency $\omega$ is chosen to be situated in the domain of the toroidicity-induced gap, so that the constraint $\omega \ll \Omega_{\rm g}$ is satisfied. The perturbation model is described in Section~\ref{sec:model}, results are presented in Section~\ref{sec:results}, and we conclude with a discussion in Section~\ref{sec:summary}. Theoretical analyses are presented in the Appendix.

% =============================================================================
\section{Model}
\label{sec:model}

The trajectories and topology of fast ion GC orbits are analyzed using the code {\tt ORBIT} \cite{White84}. We consider fast deuterons with kinetic energies around $K = 400\,{\rm keV}$ and velocity pitch $v_\parallel/v = \sin(\pi/4)$. The magnetic geometry is based on a toroidally symmetric JT-60U plasma with the same parameters as in Ref.~\onlinecite{Bierwage18}. The magnetic axis is located at major radius $R_0 = 3.4\,{\rm m}$ and height $z_0 = 0.2\,{\rm m}$. The central field strength is $B_0 = |{\bm B}_\eq(R_0,z_0)| = 1.16\,{\rm T}$ and the plasma current is $I_{\rm p} = 0.57\,{\rm MA}$. Field and current both flow in the $+\zeta$ direction, so that both $\Psi$ and $\Psi_{\rm P}$ increase monotonically from the center ($\Psi_0 = \Psi_{\rm P,0} = 0$) to the edge of the plasma ($\Psi_{\rm edge} = q_{\rm edge}\Psi_{\rm P,edge} > 0$). We use SI units in this section.

We apply an electromagnetic perturbation ${\bm E} = -\nablab\Phi - \partial_t\delta{\bm A}$ that causes displacements of the form
\begin{equation}
{\bm \xi}(\psi_{\rm P},\vartheta,\zeta,t) = \xi_0^\Psi \sum\limits_m \hat{\bm\xi}_m(\psi_{\rm P}) \sin\left(\Theta_m(\vartheta,\zeta,t)\right),
\label{eq:mode_xi}
\end{equation}

\noindent with constant amplitude $\xi_0$ and time-dependent phase
\begin{equation}
\Theta_m(\vartheta,\zeta,t) = n\zeta - m\vartheta - \omega t + \Theta_{0,m}.
\label{eq:mode_phase}
\end{equation}

\noindent The oscillation frequency $\omega = 2\pi\nu$ is fixed and has units of $[{\rm rad/s}]$. The spatial structure of the perturbation is determined by a single toroidal harmonic with mode number $n$, a set of poloidal harmonics with mode numbers $m$, and a radial profile described by the parametric model
\begin{equation}
\hat{\xi}^\Psi_m(\psi_{\rm P}) = \exp\left(-(\hat{r}(\psi_{\rm P}) - \hat{r}_{0,m})^2/\hat{r}_{{\rm w},m}^2\right).
\label{eq:xi_r}
\end{equation}

\begin{figure}[tbp]
\centering%\vspace{-0.2cm}
\includegraphics[width=0.48\textwidth]{\figures/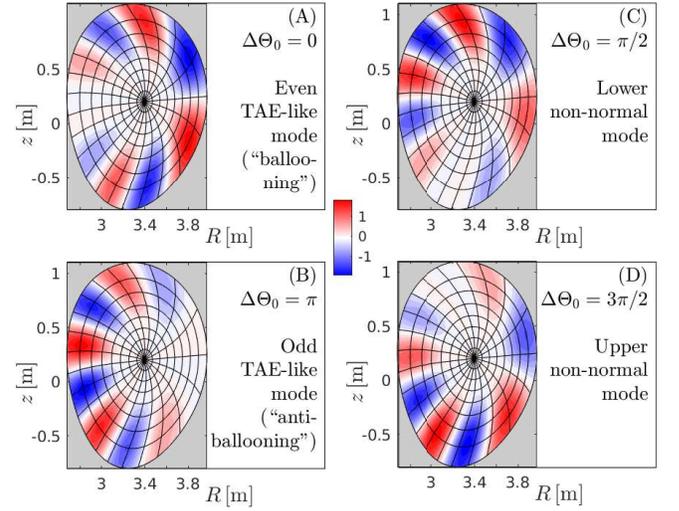}
\caption{Mode structure of the displacement $\xi^\Psi(R,z)$ modeled by Eq.~(\ref{eq:mode_xi}) for the parameters of the four cases (A)--(D) in Tables~\protect\ref{tab:parm_prof} and \protect\ref{tab:parm_pol}. Boozer coordinate lines are drawn black. Note that only the inner 90\% of the flux space are shown.}
\label{fig:02}%
\end{figure}

\noindent The minor radial coordinate $\hat{r}(\psi_{\rm P}) = r/a \in [0,1]$ is plotted in Fig.~\ref{fig:02}(a) and is approximately equal to the square root of the normalized toroidal flux $\psi = \Psi/\Psi_{\rm edge}$. The displacement vector ${\bm \xi}$ has units of length (meters) and $\hat{\xi}^\Psi_m = \hat{\bm \xi}_m\cdot\nablab\Psi \in [-1,+1]$ in Eq.~(\ref{eq:xi_r}) is dimensionless, so the amplitude factor $\xi_0^\Psi$ in Eq.~(\ref{eq:mode_xi}) has the units of $\Psi$: $[{\rm T}\cdot{\rm m}^2/{\rm rad}] = [{\rm V\cdot s}/{\rm rad}]$. The electric potential $\Phi$ of the perturbation is expanded in the same form (\ref{eq:mode_xi}) as the displacement vector ${\bm \xi}$. With the displacement vector defined as $\delta{\bm B} = \nablab\times({\bm \xi}\times{\bm B}_\eq)$, Faraday's law $\partial_t\delta{\bm B} = -\nablab\times{\bm E}$ yields the equality
\begin{equation}
\frac{\Phi_0 \hat{\Phi}_m}{(gq + I)\omega} = -\frac{\xi_0^\Psi \hat{\xi}_m^\Psi}{(mg + nI) q},
\label{eq:epot}
\end{equation}

\noindent where $\Phi_0$ has units of $[{\rm T}\cdot{\rm m}^2/{\rm s}] = [{\rm V}]$.

We assume that the perturbation has the form of an ideal incompressible electromagnetic flute mode with $E_\parallel = ({\bm B}_\eq + \delta{\bm B})\cdot{\bm E} = 0$ and $\delta{\bm B} = \nablab\times\delta{\bm A} = \nablab\alpha\times{\bm B}_\eq$. The condition $E_\parallel = 0$ means that
\begin{align}
({\bm B}_\eq + {\cgray \delta{\bm B}})\cdot\nablab\Phi &= \left(B^2_\eq + {\bm B}_\eq\cdot\nablab\times(\alpha{\bm B}_\eq)\right)\partial_t\alpha \nonumber
\\
&= B_\eq^2 ( 1 + {\cgray \alpha\mu_0 J_\parallel/B_\eq})\partial_t\alpha.
\label{eq:ideal_exact}
\end{align}

\noindent The parallel current density is $\mu_0 J_\parallel = {\bm b}\cdot(\nablab\times{\bm B}_\eq)$ with $\hat{\bm b} \equiv {\bm B}_\eq/B_\eq$ (see Eq.~(\ref{eq:jpar}) of Appendix~\ref{apdx:theory_b}). In the plasma core, we have $\mu_0 J_\parallel/B_\eq \approx (gI' - Ig' - gq\partial_\vartheta\beta_*)/(gq) \lesssim  0.2\,{\rm m}^{-1}$ (Fig.~\ref{fig:01}). Ignoring the nonlinear fluctuation terms, printed gray in Eq.~(\ref{eq:ideal_exact}), we relax $E_\parallel = 0$ to $E_\parallel \approx 0$ with
\begin{equation}
\frac{{\bm B}_\eq\cdot\nablab\Phi}{B_\eq^2} = \partial_t\alpha \quad \Leftrightarrow \quad \frac{E_\parallel}{E_\perp} = \O\left(\frac{\alpha^2 \omega}{v_{\rm E0}}\frac{\mu_0 J_\parallel}{B_\eq}\right).
\label{eq:ideal_lin}
\end{equation}

\noindent From the contravariant form of ${\bm B}_\eq$ in Eq.~(\ref{eq:b_flux}), it follows that $q\J {\bm B}_\eq\cdot\nablab = \partial_\vartheta + q\partial_\zeta$. In Boozer coordinates, the Jacobian $\J = [\nablab\Psi\cdot(\nablab\vartheta\times\nablab\zeta)]^{-1}$ satisfies $q B^2_\eq \J = g q + I$. Expanding $\Phi$ like ${\bm \xi}$ in Eq.~(\ref{eq:mode_xi}), the derivatives in Eq.~(\ref{eq:ideal_lin}) imply that $\alpha$ is expanded in cosines as
\begin{equation}
\alpha(\psi_{\rm P},\vartheta,\zeta,t) = \alpha_0 \sum\limits_m \hat\alpha_m(\psi_{\rm P}) \cos\left(\Theta_m(t)\right).
\label{eq:mode_alpha}
\end{equation}

\noindent The amplitudes of the Fourier components $\alpha_0$, $\Phi_0$ and $\xi^\Psi_0$ are then related as
\begin{equation}
\alpha_0\hat{\alpha}_m = \frac{(nq - m)}{(gq + I)} \frac{\Phi_0\hat{\Phi}_m}{\omega} = -\frac{(nq - m)}{(mg + nI)} \frac{\xi_0^\Psi\hat{\xi}_m^\Psi}{q}.
\label{eq:ideal}
\end{equation}

\noindent Note that the ideal MHD constraint (\ref{eq:ideal}) is not essential for this work. Omission of the magnetic component $\alpha$ of the perturbation merely alters the form of the orbit contours in Poincar\'{e} plots (see Fig.~\ref{fig:16} below).

\begin{table}[tbp]%\vspace{-0.2cm}
\caption{Mode parameters for Eqs.~(\protect\ref{eq:mode_xi}) and (\protect\ref{eq:xi_r}). All poloidal harmonics $m$ are taken to have the same radial profile.}
\begin{ruledtabular}
\begin{tabular}{c|c|c|c|c}
%\hline\hline
$\nu = \omega/(2\pi)$ & $n$ & $\xi_0^\Psi/B_0$ & $\hat{r}_{0,m}$ & $\hat{r}_{{\rm w},m}$ \\
\hline $47.4\,{\rm kHz}$ & $2$ & $6\times 10^{-3}\,{\rm m}^2/{\rm rad}$ & $0.8$ & $0.15$ \\
%\hline\hline
\end{tabular}
\label{tab:parm_prof}
\end{ruledtabular}
\end{table}

\begin{table}[tbp]\begin{ruledtabular}
\caption{Poloidal Fourier harmonics $m$ and their phases $\Theta_{0,m}$ characterizing the $\vartheta$-dependence of the perturbation in Eq.~(\protect\ref{eq:mode_xi}).}
\begin{tabular}{c|c|c|c|c|c}
%\hline\hline
Case & $m_{\rm a}$ & $m_{\rm b}$ & $-\Theta_{0,{\rm a}}$ & $-\Theta_{0,{\rm b}}$ & $\Delta\Theta_0 = \Theta_{0,{\rm b}} - \Theta_{0,{\rm a}}$ \\
\hline (A) & $4$ & $5$ & $\pi/2$ & $\pi/2$ & $0$ \\
(B) & $4$ & $5$ & $3\pi/2$ & $3\pi/2 - \pi$ & $\pi$ \\
(C) & $4$ & $5$ & $\pi$ & $\pi - \pi/2$ & $\pi/2$ \\
(D) & $4$ & $5$ & $0$ & $0 - 3\pi/2$ & $3\pi/2$ \\
%\hline\hline
\end{tabular}
\end{ruledtabular}
\label{tab:parm_pol}
\end{table}

The particles in our simulations are subject to a mode with toroidal mode number $n = 2$ and frequency $\nu = \omega/(2\pi) = 47.4\,{\rm kHz} \approx 5\times 10^{-3}\Omega_{\rm g0}/(2\pi)$. These and the values of the parameters for the radial profile defined in Eq.~(\ref{eq:xi_r}) are summarized in Table~\ref{tab:parm_prof}.

For the $\vartheta$-dependence of the mode structure, we consider four cases labeled (A)--(D) whose parameters are shown in Table~\ref{tab:parm_pol}. All cases have the same pair of poloidal harmonics $m_{\rm a} = 4$ and $m_{\rm b} = 5$. We only vary the phases $\Theta_{0,m}$. The absolute values of $\Theta_{0,m}$ are chosen such that the elliptic point of the resonances will appear at $\vartheta = 0$. The main parameter of interest is the phase difference $\Delta\Theta_0 = \Theta_{0,{\rm b}} - \Theta_{0,{\rm a}}$, which determines the poloidal location of a mode's peak through the constructive interference of its two components $m_{\rm a}$ and $m_{\rm b}$ as shown in Fig.~\ref{fig:02}. The interference patterns can be readily understood by writing the two superimposed waves as
\begin{align}
&\exp\left(i\Theta_{0,{\rm b}} - i m_{\rm b}\vartheta\right) + \exp\left(i\Theta_{0,{\rm a}} - i m_{\rm a}\vartheta\right)
\label{eq:beat}
\\
&= \underbrace{2 \cos\left(\frac{\Delta\Theta_0 - \vartheta\Delta m}{2}\right)}\limits_{\rm beat\, envelope} \underbrace{\exp\left(i\overline{\Theta}_0 - i\frac{m_{\rm b}+m_{\rm a}}{2}\vartheta\right)}\limits_{\rm base\, oscillation} \nonumber
\end{align}

\noindent with $\Delta m = m_{\rm b} - m_{\rm a} = 1$ and $\overline{\Theta}_0 = (\Theta_{0,{\rm b}} + \Theta_{0,{\rm a}})/2$. One can readily see that $\vartheta = \Delta\Theta_0/\Delta m$ identifies the location of maximal constructive interference, whereas destructive interference occurs at $\vartheta = (\Delta\Theta_0 - \pi)/\Delta m$, independently of the position along the toroidal angle $\zeta$.

Cases (A) and (B) have $\Delta\Theta_0 = 0$ and $\pi$. They resemble the structure of toroidicity-induced Alfv\'{e}n eigenmodes (TAE) \cite{Cheng85} with (A) even and (B) odd parity, which peak on the plasma's outboard and inboard side, respectively. Cases (C) and (D) have $\Delta\Theta_0 = \pi/2$ and $3\pi/2$, so that their peaks are located in the upper and lower part of the plasma, respectively. We refer to them as nonnormal modes, since we are not aware of any linear eigenmodes in a tokamak that have such a structure. However, such structures may form at least transiently in the course of nonlinear plasma dynamics, or they may be imposed by external forcing.

Since our mode structures are defined in terms of Fourier harmonics in the nonorthogonal coordinates $\Psi_{\rm P}$ and $\vartheta$, the ${\bm E}\times{\bm B}$ vortices appear sheared when plotted in physical space as in Fig.~\ref{fig:02}. We consider only orbits in the region $0.1 \lesssim \psi_{\rm P} \lesssim 0.95$, where the Boozer coordinates' nonorthogonality amounts to $q\beta_* \lesssim 0.2\,{\rm m^{-1}}$ as one can see in Fig.~\ref{fig:01}(b).

We prescribe the perturbation amplitude via the displacement $\xi_0^\Psi/B_0 = 6\times 10^{-3}\,{\rm m}^2/{\rm rad}$ (units of {\tt ORBIT}). This corresponds to $|\alpha_0| \approx |nq-m|\xi^\Psi_0/(m q R_0 B_0) \lesssim 2\times 10^{-4}\,{\rm m}/{\rm rad}$ and $\Phi_0 \approx \xi^\Psi_0 \omega/m \approx 0.5\,{\rm kV}$ per Fourier harmonic. The potential variations of $m_{\rm a}$ and $m_{\rm b}$ combined have then a magnitude of up to $1\,{\rm kV}$ and a maximal potential difference of $|\Delta\Phi| \sim 2\,{\rm kV}$. The corresponding electric drift velocity is ${\rm max}|{\bm v}_{\rm E}| \sim |\Delta\Phi|m/(r B_0) \approx 10^4\,{\rm m/s}$, which is about $0.2\%$ of the on-axis Alfv\'{e}n velocity $v_{\rm A0} \approx 4.3\times 10^6\,{\rm m/s}$. This is a large but realistic value. For comparison, we have observed velocities up to ${\rm max}|{\bm v}_{\rm E}(n=2)|/v_{\rm A0} \approx 0.6\%$ in simulations of Abrupt Large-amplitude Events (ALE) in the same JT-60U plasma driven by negative-ion-based neutral beams \cite{Bierwage18,Bierwage22c}.

\begin{figure}[tb]
\centering\vspace{-0.25cm}
\includegraphics[width=0.48\textwidth]{\figures/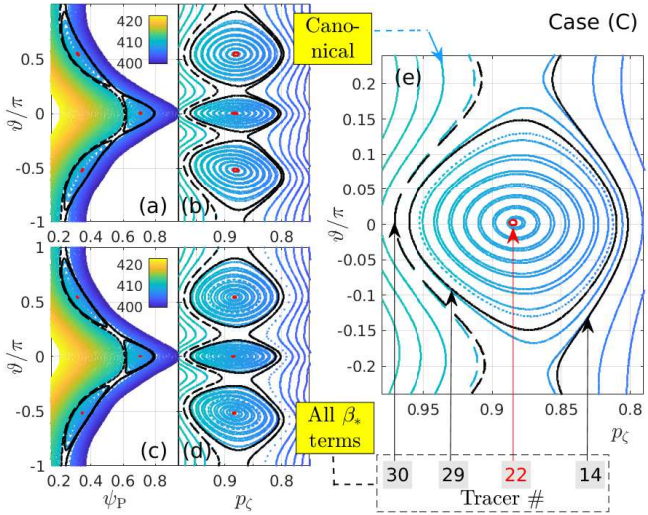}\vspace{-0.1cm}
\caption{Comparison of the resonance structures for (a,b) the canonical system with $\beta_* \rightarrow 0$, and (c,d) the complete system with all $\beta_*$ terms retained. The time step used in these calculations was reduced by a factor 10 to $\Delta t = 0.048\times\Omega_{\rm g0}^{-1}$ ($1/4000$ of a toroidal transit time), where the numerical accuracy becomes independent of $\Delta t$ (Fig.~\protect\ref{fig:04}). Poincar\'{e} plots are shown for the nonnormal mode case (C) in two sets of polar coordinates: $(\psi_{\rm P},\vartheta)$ and $(p_\zeta,\vartheta)$. The $p_\zeta$ axis has been inverted for better comparability with the $\psi_{\rm P}$ axis. Tracer particles are loaded uniformly in the interval $0.2 \leq \psi_{\rm P} \leq 0.95$ at $\vartheta = 0$ and with identical values of $E'_0$. The colors represent the instantaneous total energy $\E = K + \Phi$ in ${\rm keV}$, starting from $K_0 = 400\,{\rm keV}$ at $\psi_{\rm P} = 0.95$ and rising to about $423\,{\rm keV}$ at $\psi_{\rm P} = 0.2$. In panel (e), the bluish contours of (b) are overlaid with the four Poincar\'{e} contours (red, black) associated with tracers \#14, 22, 29, 30 from (d).}
\label{fig:03}%
\end{figure}

% =============================================================================
\section{Results}
\label{sec:results}

{\tt ORBIT} uses the 4th-order Runge-Kutta (RK4) scheme whose accuracy improves with smaller step size. Most of the results presented in this section were obtained with time steps of size $\Delta t = 0.48\times\Omega_{\rm g0}^{-1}$ ($1/400$'th part of a toroidal transit time), which is sufficient for the $30\,{\rm ms}$ (millisecond) period of physical time that we simulated.

The Hamiltonian $\H$ on which {\tt ORBIT} is based does not include the ponderomotive potential of Eq.~(\ref{eq:h}), so the total energy of a particle (normalized as in Section~\ref{sec:intro}) is given by
\begin{equation}
\E = K + \Phi \quad {\rm with} \quad K = \rho_\parallel^2 B^2/2 + \mu B.
\end{equation}

\noindent Since we consider fairly energetic deuterons with $K = 400\,{\rm keV}$, the potential difference $|\Delta\Phi| \sim 2\,{\rm kV}$ is tiny in comparison, and the difference between $K$ and $E$ is minute as well. Even for resonant particles, the energy exchange will be seen to reach only about $7\,{\rm keV}$.

Poincar\'{e} plots will be shown in two sets of polar coordinates: $(\psi_{\rm P},\vartheta)$ and $(p_\zeta,\vartheta)$, where
\begin{equation}
p_\zeta = \frac{P_\zeta}{\Psi_{\rm P,edge}} = \frac{\P_\zeta - \alpha g}{\Psi_{\rm P,edge}} = \frac{g \rho_\parallel}{\Psi_{\rm P,edge}} - \psi_{\rm P}
\end{equation}

\noindent is the normalized canonical toroidal angular momentum for the {\it unperturbed} magnetic field. We will be looking at the conservation of the rotating frame energy
\begin{equation}
\E' = \E - \omega\P_\zeta/n = \overbrace{\underbrace{K - \omega P_\zeta/n}\limits_{E'_0} + \Phi}\limits^{E'} - \omega\alpha g/n,
\label{eq:erot_all}
\end{equation}

\noindent where $E'_0$ is the initial value and $E' = \E' + \omega\alpha g/n$ excludes the magnetic fluctuation. For background information, see Appendix~\ref{apdx:theory_erot}.

\begin{figure}[tb]
\centering\vspace{-0.25cm}
\includegraphics[width=0.48\textwidth]{\figures/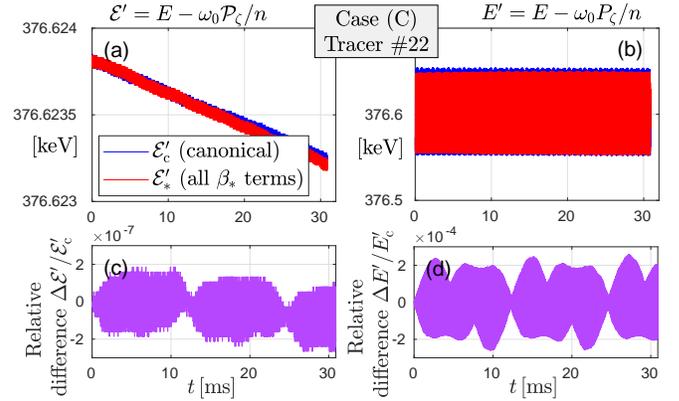}\vspace{-0.15cm}
\caption{Quality of energy conservation in the simulation of the nonnormal mode case (C). As a representative example, data are shown for tracer \#22 near the O-point of the resonance of Fig.~\protect\ref{fig:03}. Panels (a) and (b) show time traces of $\E'$ and $E'$ as defined in Eq.~(\protect\ref{eq:erot_all}) for the canonical system without $\beta_*$ (red) and the complete system with all $\beta_*$ terms retained (blue). Panels (c) and (d) show the respective relative difference between the blue and red curves. These results are converged with respect to RK4 time step $\Delta t$, so the slow drift of $\E'$ must have another cause.}
\label{fig:04}%
\end{figure}

\vspace{-0.1cm}
% -----------------------------------------------------------------------------
\subsection{Comparison of simulations using the canonical (without $\beta_*$) and the complete Lagrangian (with $\beta_*$)}
\label{sec:results_bench}

Figure~\ref{fig:03} shows that the simulation of case (C) with the complete GC phase-space Lagrangian including $\beta_*$ (a,b) yields a resonance structure that agrees with the canonical one (c,d) to a degree that seems reasonable for most practical purposes. All tracer particles were initialized with the same value of $E'_0$, which means that all Poincar\'{e} contours in our simulation scenario must be closed curves that do not overlap anywhere even after projection into the $(p_\zeta,\vartheta)$ plane. This is indeed the case in both simulations: the GC orbit topology in Fig.~\ref{fig:03} consists of distinct invariant surfaces.

\begin{figure}[tbp]
\centering%\vspace{-0.2cm}
\includegraphics[width=0.48\textwidth]{\figures/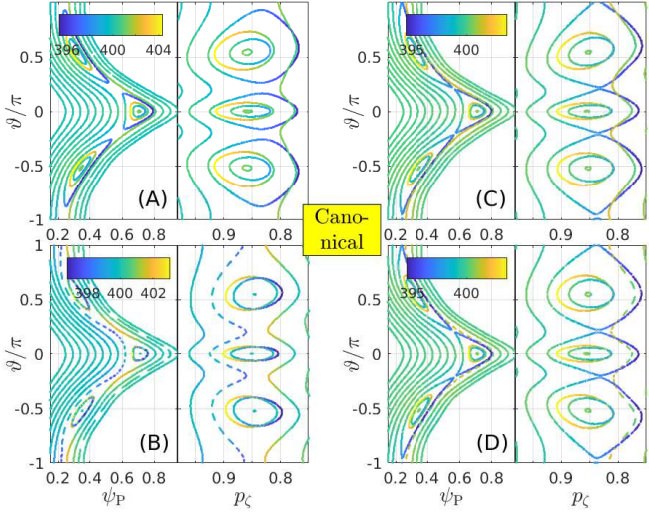}%\vspace{-0.1cm}
\caption{Resonance structures for the canonical system, with $\beta_* \rightarrow 0$, in our four cases (A)--(D). Results for normal modes are shown on the left (A, B) and results for nonnormal modes on the right (C, D). These Poincar\'{e} plots were prepared similarly to those in Fig.~\protect\ref{fig:03}, except that all tracer particles were initialized with the same kinetic energy $K_0 = 400\,{\rm keV}$. The colors represent again the instantaneous total energy $\E = K + \Phi$ in ${\rm keV}$.}
\label{fig:05}%
\end{figure}

The enlarged overlay of the results in Fig.~\ref{fig:03}(e) shows that the location of the O-point (red) is shifted slightly inward in $\psi_{\rm P}$ (towards larger $p_\zeta$). Tolerably small differences can also be seen in the contours around the separatrix (black). Note that there is always one point on each contour where the results match exactly, simply because both simulations (with and without $\beta_*$) use the same initial conditions for all tracer particles.

Figure~\ref{fig:04}(a) shows that the simulation of the complete system conserves the rotating frame energy $\enr' = \E - \omega\P_\zeta/n$ with high accuracy. With a relative difference $\Delta \E'/\E'_{\rm c} = (\E'_* - \E'_{\rm c})/\E'_{\rm c}$ on the order of $10^{-7}$ in Fig.~\ref{fig:04}(c), the result of the complete system with $\beta_*$, here denoted as $\E'_*$, is essentially identical to that of the canonical simulation, written $\E'_{\rm c}$. The results shown in Fig.~\ref{fig:04}(a) are numerically converged with respect to the RK4 time step $\Delta t$, so that the (very slow) energy drift that is visible in both simulations (canonical and noncanonical) must be due to reasons other than temporal resolution. One suspect is the spline representation of the magnetic field, perhaps in combination with modulations associated with the up-down asymmetry of our plasma \cite{Matsuyama17}.

\begin{figure}[tbp]
\centering%\vspace{-0.1cm}
\includegraphics[width=0.48\textwidth]{\figures/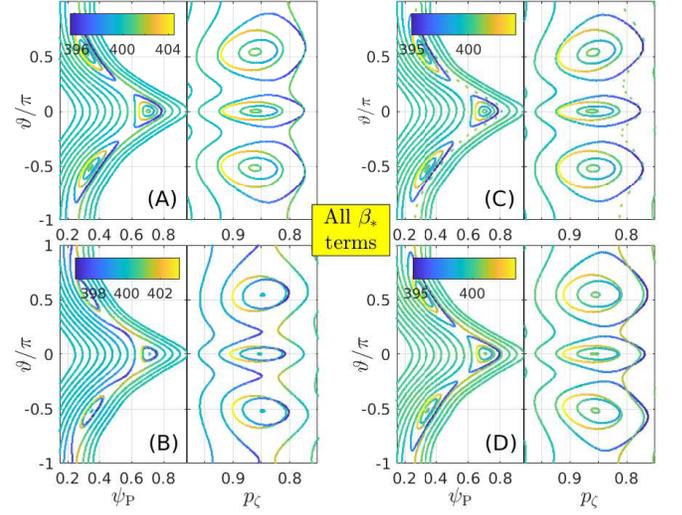}%\vspace{-0.1cm}
\caption{Resonance structures for the complete system with all $\beta_*$ terms retained. Showing cases (A)--(D), arranged as in Fig.~\protect\ref{fig:05}.}
\label{fig:06}%
\end{figure}

\begin{figure}[tbp]
\centering%\vspace{-0.2cm}
\includegraphics[width=0.48\textwidth]{\figures/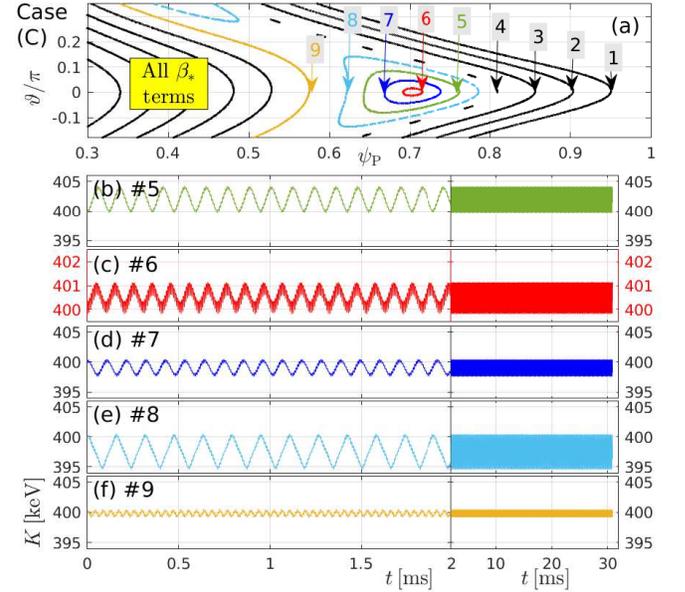}%\vspace{-0.1cm}
\caption{Time traces of the kinetic energies $K(t)$ of tracer particles \#5--\#9 in the nonnormal mode case (C) simulated for the complete system with all $\beta_*$ terms retained.  Orbit \#6 (red) is trapped close to the O-point of the resonance, so we have chosen a smaller energy range $(401 \pm 1.5)\,{\rm keV}$ for panel (c). The other panels have axes limits at $(400 \pm 6)\,{\rm keV}$.}
\label{fig:07}%
\end{figure}

\begin{figure}[tbp]
\centering%\vspace{-0.25cm}
\includegraphics[width=0.48\textwidth]{\figures/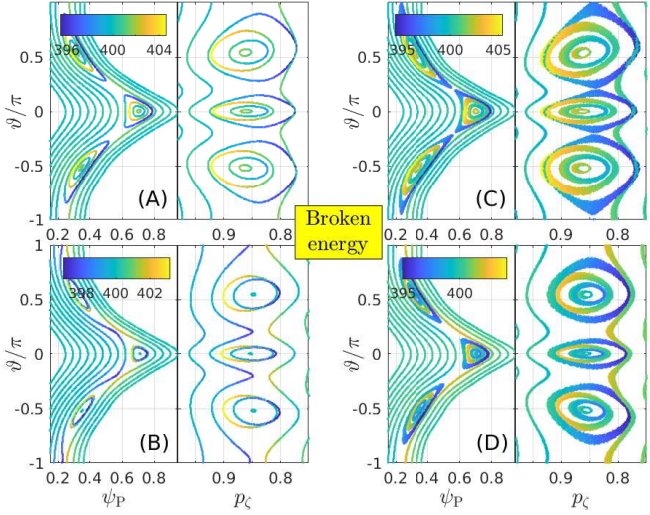}%\vspace{-0.1cm}
\caption{Resonance structures for the broken system violating energy conservation due to the omission of terms containing $\alpha'_\vartheta\beta_*$ and $\alpha'_\zeta\beta_*$. Showing cases (A)--(D), arranged as in Fig.~\protect\ref{fig:05}.}
\label{fig:08}%
\end{figure}

\begin{figure}[tbp]
\centering%\vspace{-0.2cm}
\includegraphics[width=0.48\textwidth]{\figures/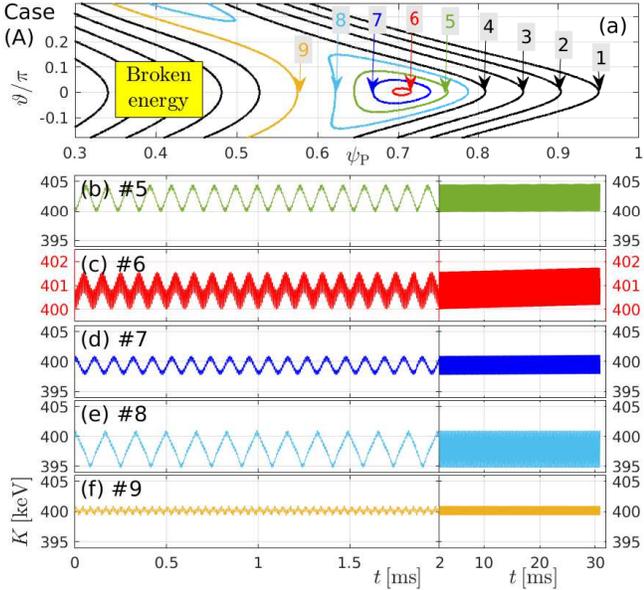}
\caption{Time traces of the kinetic energies $K(t)$ of tracer particles \#5--\#9 in the normal mode case (A) simulated for the broken system violating energy conservation. Arranged as Fig.~\protect\ref{fig:07}. Similar results (not shown) are obtained in case (B).}
\label{fig:09}%
\end{figure}

\begin{figure}[tbp]
\centering%\vspace{-0.25cm}
\includegraphics[width=0.48\textwidth]{\figures/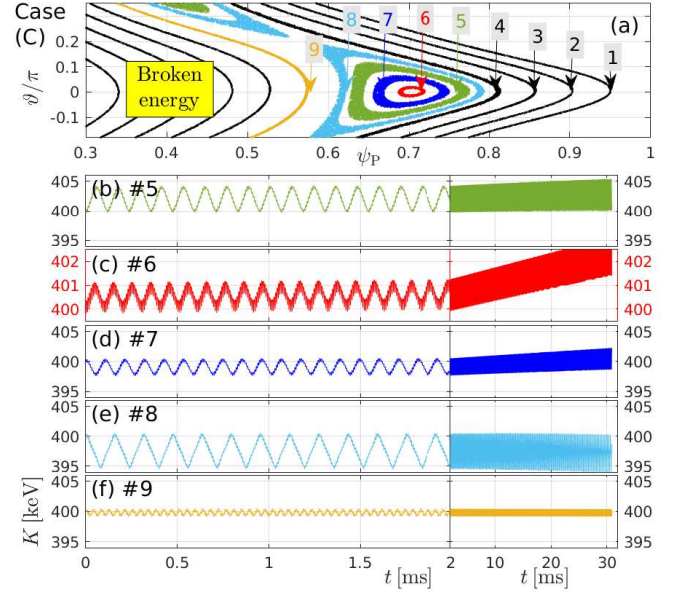}
\caption{Time traces of the kinetic energies $K(t)$ of tracer particles \#5--\#9 in the nonnormal mode case (C) simulated for the broken system violating energy conservation. Arranged as Fig.~\protect\ref{fig:07}. Similar results (not shown) are obtained in case (D), except that the direction of secular acceleration and displacement is reversed, as can be seen in Fig.~\protect\ref{fig:11}(D) for tracer \#6.}%\vspace{-0.0cm}
\label{fig:10}%
\end{figure}

Since our Poincar\'{e} plots were prepared using the modified momentum coordinate $P_\zeta = \P_\zeta - g\alpha$ (without the magnetic perturbation), we show for completeness also the time traces of $E' = \E - \omega P_\zeta/n$ in Fig.~\ref{fig:04}(b). One can see that $E'$ is conserved on average. The relative difference $\Delta E'/E'_{\rm c} = (E'_* - E'_{\rm c})/E_{\rm c}'$ between the canonical and noncanonical simulations is shown in Fig.~\ref{fig:04}(d), where it can be seen to reach amplitudes amounting to $\omega|\P_\zeta - P_\zeta|/(n|E'_{\rm c}|) \approx \omega g\alpha_0/(n K) \approx {\rm few}\times 10^{-4}$ within about $1\,{\rm ms}$. Subsequently, $\Delta E'(t)$ is subject to multi-harmonic pulsations. The most prominent cycle is about $12\,{\rm ms}$ long, where $\Delta E'(t)$ nearly vanishes.

This $12\,{\rm ms}$ cycle is also seen in $\Delta\E'(t)$ in Fig.~\ref{fig:04}(c). It indicates that the GC orbit in the noncanonical simulation slowly drifts with respect to results of the canonical simulation, and that the phase shift reaches $2\pi$ after about $12\,{\rm ms}$, while staying on the same orbit contour; i.e., on the same $\E' = {\rm const}$.\ surface. This is consistent with the theory, which says that retention of $\beta_*$ corresponds merely to a renormalization of time in the unperturbed axisymmetric case ($\alpha = \Phi = 0$) (see p.~78 of Ref.~\onlinecite{WhiteTokBook3}). The pulsations in Fig.~\ref{fig:04}(c,d) suggest that this is true even for the perturbed system with nonzero $\Phi$ and $\alpha$.

Figures~\ref{fig:05} and \ref{fig:06} show, respectively, the resonance structures of the canonical and noncanonical simulations for all four cases (A)--(D). Here, all tracer particles were initialized with the same kinetic energy $K_0 = 400\,{\rm keV}$, so that some Poincar\'{e} contours overlap when projected into the $(p_\zeta,\vartheta)$ plane. The contours are distinct in energy $\E = K + \Phi$, as can be seen from the colors in these plots. The island width is smallest in case (B) because our co-passing deuteron orbits are shifted radially outward in $R$, where the mode amplitude is small in case (B).

Finally, Fig.~\ref{fig:07} shows time traces of the kinetic energy $K = \E - \Phi$ for five tracer particles that are located inside or nearby the resonant island. For the entire duration of the simulation ($30\,{\rm ms}$), energy is well-conserved on average up to the level of accuracy of our numerical scheme.

\begin{figure}[tbp]
\centering%\vspace{-0.05cm}
\includegraphics[width=0.48\textwidth]{\figures/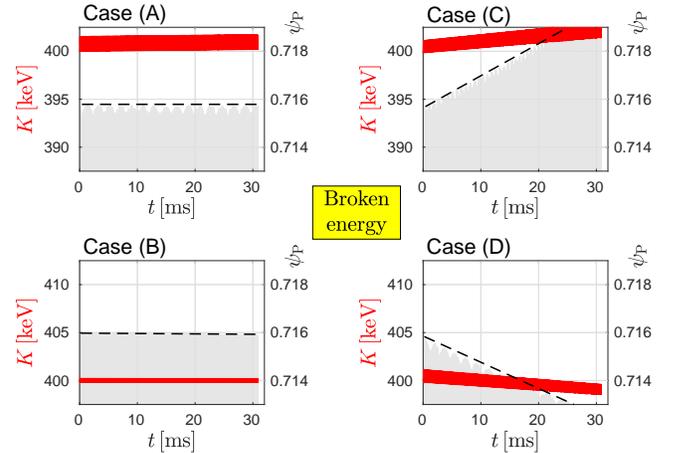}%\vspace{-0.1cm}
\caption{Time traces of the kinetic energy $K(t)$ (red) and radial position in terms of normalized poloidal flux $\psi_{\rm P}(t)$ (gray) for tracer \#6 near the O-point of the resonance in cases (A)--(D) simulated for the broken system violating energy conservation. Note that $\psi_{\rm P}(t)$ varies largely ($0.224...0.716$) during each poloidal transit, so only the upper rim of the rapidly oscillating curve is shown and marked by a black dashed line.}
\label{fig:11}%
\end{figure}

\begin{figure}[tbp]
\centering\vspace{-0.25cm}
\includegraphics[width=0.48\textwidth]{\figures/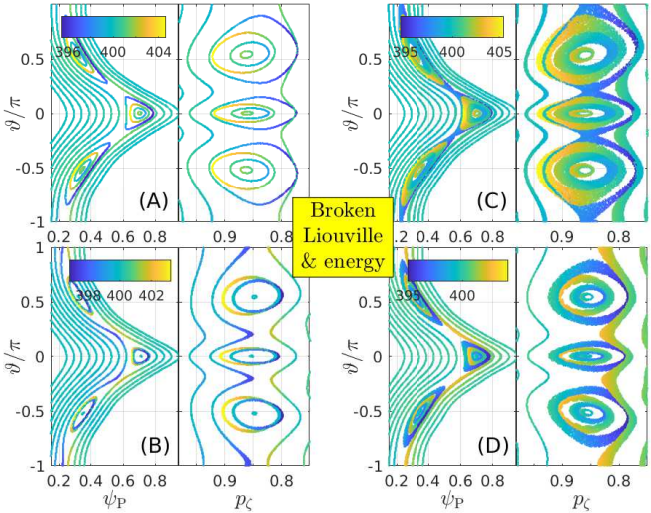}\vspace{-0.1cm}
\caption{Resonance structures for the partial system where both the Liouville theorem and energy conservation are violated due to omission a small term containing $\alpha'_\zeta\beta_*$. Showing cases (A)--(D), arranged as in Fig.~\protect\ref{fig:05}.}
\label{fig:12}%
\end{figure}

\begin{figure}[tbp]
\centering%\vspace{-0.2cm}
\includegraphics[width=0.48\textwidth]{\figures/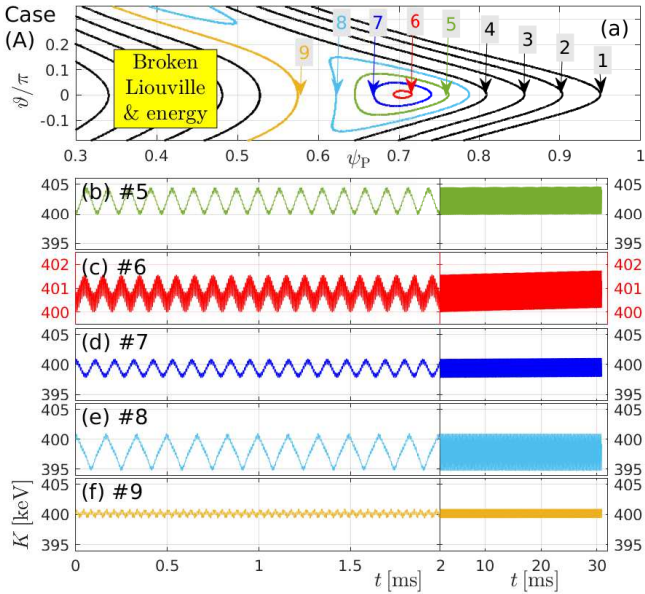}\vspace{-0.1cm}
\caption{Time traces of the kinetic energies $K(t)$ of tracer particles \#5--\#9 in the normal mode case (A) simulated for the broken system violating both the Liouville theorem and energy conservation. Arranged as Fig.~\protect\ref{fig:07}. Case (B) is similar (not shown).}
\label{fig:13}%
\end{figure}

\begin{figure}[tbp]
\centering\vspace{-0.25cm}
\includegraphics[width=0.48\textwidth]{\figures/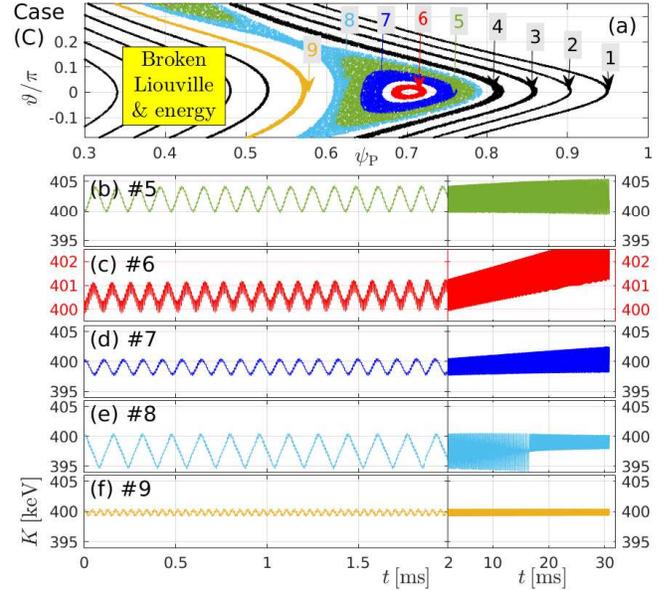}\vspace{-0.1cm}
\caption{Time traces of the kinetic energies $K(t)$ of tracer particles \#5--\#9 in the nonnormal mode case (C) simulated for the broken system violating both the Liouville theorem and energy conservation. Arranged as Fig.~\protect\ref{fig:07}. Case (D) is similar (not shown), except that the direction of secular acceleration and displacement is reversed, as can be seen in Fig.~\protect\ref{fig:15}(D) for tracer \#6.}\vspace{-0.1cm}
\label{fig:14}%
\end{figure}

\begin{figure}[tbp]
\centering%\vspace{-0.05cm}
\includegraphics[width=0.48\textwidth]{\figures/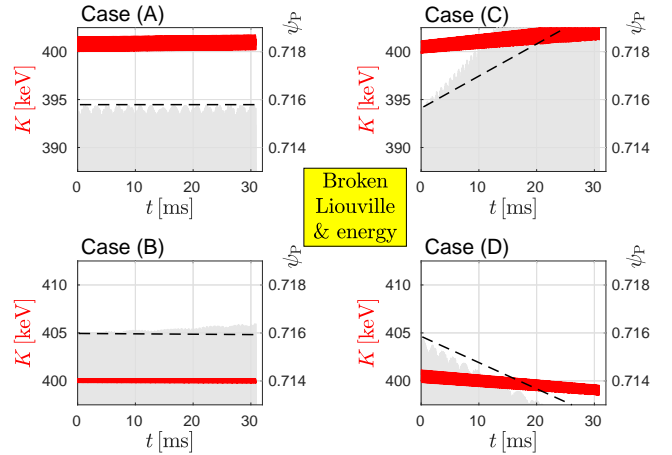}\vspace{-0.1cm}
\caption{Time traces of the kinetic energy $K(t)$ (red) and radial position in terms of normalized poloidal flux $\psi_{\rm P}(t)$ (gray) for tracer \#6 near the O-point of the resonance in cases (A)--(D) simulated for the broken system violating both the Liouville theorem and energy conservation. Note that $\psi_{\rm P}(t)$ varies largely ($0.224...0.716$) during each poloidal transit, so only the upper rim of the rapidly oscillating curve is shown. The black dashed lines are the same as in Fig.~\protect\ref{fig:11} in order to highlight the different rates of secular radial drift.}
\label{fig:15}%
\end{figure}

% -----------------------------------------------------------------------------
\subsection{Breaking the conservative character by omitting small terms $\alpha'_\vartheta\beta_*$ and/or $\alpha'_\zeta\beta_*$}
\label{sec:results_break}

Appendices~\ref{apdx:theory_liouville} and \ref{apdx:theory_erot} contain an analysis of the equations of motion, showing that the noncanonical system that includes $\beta_*$ maintains two important conservative properties of the Hamiltonian system, which are theoretically guaranteed in the canonical case: (i) phase space conservation, and (ii) conservation of the rotating frame energy $\E'$. The results reported in the previous Section~\ref{sec:results_bench} are consistent with that analysis.

In order to see the contrast, it is interesting to inspect a few counter-examples, where energy or phase space conservation (Liouville theorem) is broken. According to the derivations in Appendices~\ref{apdx:theory_liouville} and \ref{apdx:theory_erot}, one way to achieve this is to omit small terms containing $\alpha'_\vartheta\beta_*$ and $\alpha'_\zeta\beta_*$ that appear in the equations of motion (\ref{eq:eom_matrix}) and are highlighted in red color in Eq.~(\ref{eq:fck}). Here, we use the short-hand notation $\alpha'_\eta \equiv \partial_\eta\alpha$ for partial derivatives. The omission of both terms violates energy conservation, while the Liouville theorem is still satisfied. The results for this scenario are summarized in Figs.~\ref{fig:08}--\ref{fig:11}. If only one term is omitted, both the Liouville theorem and energy conservation broken. We have arbitrarily chosen to omit $\alpha'_\zeta\beta_*$ and the results are summarized in Figs.~\ref{fig:12}--\ref{fig:15}. Both cases show secular changes in position and energy, which are numerically robust. Further reduction of the time step of the RK4 solver has no effect.

The secular acceleration and displacement is largest for particles close to the O-point of the resonance, like our tracer \#6 in Figs.~\ref{fig:09}--\ref{fig:11} and \ref{fig:13}--\ref{fig:15}. Moreover, the observed deviations from conservative motion are particularly large in cases (C) and (D) with nonnormal modes that peak around $\vartheta \approx \pm\pi/2$. In these cases, the effect becomes noticeable on the $3\,{\rm ms}$ time scale ($\sim 1000$ toroidal transits) and grows to the percentile level (few ${\rm keV}$) by the end of the $30\,{\rm ms}$ time window simulated here. The main difference between the cases with broken energy conservation and the cases where both phase space conservation and energy conservation are broken is that the latter scenario leads to stronger secular displacement in space. Secular acceleration is only slightly enhanced. The stronger radial drift is clearly visible in the Poincar\'{e} plot in Fig.~\ref{fig:14}(a) and in the evolution of $\psi_{\rm P}(t)$ in Fig.~\ref{fig:15}(C,D) for tracer \#6. In addition, one can see in Fig.~\ref{fig:14}(e) that tracer \#8 (light blue) becomes untrapped as it crosses the separatrix of the resonance at $t \approx 16.5\,{\rm ms}$.

From the practical point of view, these deviations from conservative motion on the $10\,{\rm ms}$ time scale may be tolerable because particle collisions tend to become important on the millisecond scale. Moreover, Alfv\'{e}nic instabilities in real plasmas rarely maintain large amplitudes and fixed frequencies for longer than a millisecond. Amplitude pulsations and frequency chirping usually occur within that time frame. Last but not least, spontaneously formed nonnormal modes --- which may enhance secular drifts as found in our cases (C) and (D) --- tend to be short-lived in real plasmas.

In any case, the deliberate breaking of conservation laws as performed in this section only served the purpose of visualizing the consequences and demonstrating the influence of the mode structure. The conservative system (with or without $\beta_*$) as implemented in {\tt ORBIT} is, of course, the preferred choice.

%\vspace{-0.2cm}
% =============================================================================
\section{Discussion}%\vspace{-0.1cm}
\label{sec:summary}

Using the guiding center (GC) orbit-following code {\tt ORBIT} in Boozer coordinates, we have analyzed the motion of energetic deuterons in a realistic setting based on the JT-60U tokamak. The particles were subject to perturbations resembling normal and nonnormal Alfv\'{e}n modes in terms of spatial structure and oscillation frequency. It was shown that the equations of motion deriving from the GC phase-space Lagrangian in {\it noncanonical} form --- which include the geometric coefficient $q\beta_* = B_{\Psi_{\rm P}} = {\bm B}_\eq\cdot\partial_{\Psi_{\rm P}}{\bm x}$ --- yield conservative dynamics just like the widely used canonical system, where all terms containing $\beta_*$ are omitted. The results can be said to be in agreement since differences in the orbit shapes and periods are small (Section~\ref{sec:results_bench}).

The numerical simulation results are corroborated by theoretical analyses in Appendices~\ref{apdx:theory_liouville} and \ref{apdx:theory_erot}, where it is proven that the complete system including all $\beta_*$ terms satisfies at least two important conservation laws of Hamiltonian flows: (i) phase space conservation as expressed by the Liouville theorem, and (ii) conservation of the effective energy $\E' = \E - \omega\P_\zeta/n$ in the frame of reference rotating with a mode that has a fixed displacement amplitude $\xi^\Psi_0$, a fixed frequency $\omega$, and a single toroidal mode number $n$.

\begin{figure}[tbp]
\centering\vspace{-0.45cm}
\includegraphics[width=0.48\textwidth]{\figures/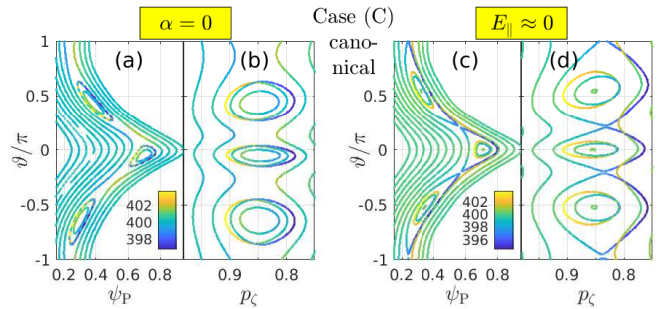}\vspace{-0.25cm}
\caption{Resonance structures for the canonical system with $\beta_* \rightarrow 0$ in the nonnormal mode case (C). Panels (a,b) show results of the simulation without magnetic perturbations ($\alpha = 0$, $E_\parallel \neq 0$). Panels (c,d) show the same results as Fig.~\protect\ref{fig:05}(C), where $\alpha$ and $\Phi$ satisfy Eq.~(\protect\ref{eq:ideal_lin}) such that $E_\parallel/E_\perp \lesssim 10^{-6} \approx 0$. The colors represent the instantaneous total energy $\E = K + \Phi$ in ${\rm keV}$.}
\label{fig:16}%
\end{figure}

In addition to these conditions, the perturbations in our simulations satisfied the constraint $E_\parallel = 0$ to a high degree of accuracy ($E_\parallel/E_\perp \lesssim 10^{-6}$) via Eq.~(\ref{eq:ideal_lin}) which enforces a correlation between the electric and magnetic fluctuations expressed in terms of $\Phi$ and $\alpha$. It should be noted, however, that this constraint is not essential. The derivations in Appendices~\ref{apdx:theory_liouville} and \ref{apdx:theory_erot} are independent of the presence or absence of $\alpha$. There are, however, some quantitative differences as illustrated in Fig.~\ref{fig:16}, where we compare Poincar\'{e} plots of the resonance in our working example for the canonical system without (left) and with (right) magnetic perturbations $\alpha$.

The mode and resonance that we analyzed here lay in a region where $q \beta_* \sim r/R_0 \lesssim 0.2$ and we saw only insignificant differences between simulations with and without $\beta_*$. However, as one can see in Fig.~\ref{fig:01}(b), $q\beta_*$ can far exceed unity in the edge region of a diverted plasma. The study of orbits that pass through the edge region with large $q\beta_*$ is of practical interest since orbits passing through the boundary and the surrounding vacuum are useful for diagnostics and can potentially damage plasma facing components of the wall. This subject is left for future work as it requires careful modeling of the plasma edge, including radial electric fields and deviations from axisymmetry in ${\bm B}_\eq$. The latter factor is also connected with the issue of handling problematic Pfirsch-Schl\"{u}ter currents that $\beta_*$ represents in Boozer coordinates as was briefly discussed in the introduction (Section~\ref{sec:intro}).

As mentioned in the introduction Section~\ref{sec:intro}, the field perturbations applied in {\tt ORBIT} automatically satisfy Maxwell's equations since they are expressed in terms of potentials $\Phi$ and ${\bm A}$. This is not necessarily the case in models using the physical fields ${\bm E}$ and ${\bm B}$, where small discrepancies in the mode structure itself can lead to unphysical secular dynamics as shown in the companion paper, Ref.~\onlinecite{Bierwage22c}.

We recommend that models and codes be tested not only using normal modes as obtained from linear eigenvalue solvers, but also using nonnormal modes like our cases (C) and (D) in Fig.~\ref{fig:02}. In Section~\ref{sec:results_break} we showed that the nonnormal modes can reveal inconsistencies in the model more clearly than normal modes do. Although our cases (A) and (B) were not exact normal modes, they capture as a relevant feature the in-out asymmetry of modes found in toroidal plasmas. In contrast, the strong up-down asymmetry of our nonnormal modes in cases (C) and (D) is out-of-phase with the modulation of $B_\eq$.

\vspace{-0.25cm}
\begin{acknowledgments}\vspace{-0.25cm}
One of the authors (A.B.) is grateful to Vin\'{i}cius Duarte  and Timur Esirkepov for helpful discussions. The workstation used for the numerical calculations reported here was funded by QST President's Strategic Grant (Creative Research). The work by A.B.\ was partially supported by MEXT as ``Program for Promoting Researches on the Supercomputer Fugaku'' (Exploration of burning plasma confinement physics, JPMXP1020200103). The work by R.B.W.\ was supported by the US Department of Energy (DOE) under contract DE-AC02-09CH11466.
\end{acknowledgments}

\vspace{-0.35cm}
\section*{Data Availability Statement}\vspace{-0.25cm}

The data that support the findings of this study are available from the corresponding author upon reasonable request.

\appendix

% =============================================================================
\section{Guiding center model}
\label{apdx:theory}

In this Appendix we revisit the theory of guiding center (GC) motion for particles with electric charge $Ze$ and mass $M$. The review by Cary \& Brizard \cite{Cary09} and the book by White \cite{WhiteTokBook3} served as guides, but our notation differs in parts. The results of this derivation corroborate the numerical results in the main part of the paper. The derivation is also used to determine suitable (small) terms for demonstratively breaking certain conservation laws in a systematic way in Section~\ref{sec:results_break}.

%-------------------------------------------------------------------------------
\subsection{Guiding center phase space Lagrangian}
\label{apdx:theory_lgc}

GC theory exploits the small parameters
\begin{equation}
\epsilon_B = \frac{\rho_{\rm g}}{L_B} \ll 1, \;\; \epsilon_\parallel = \frac{|k_\parallel v_\parallel|}{\Omega_{\rm g}} \ll 1, \;\; \epsilon_\omega = \frac{\omega}{\Omega_{\rm g}} \ll 1.
\end{equation}

\noindent Small $\epsilon_B$ means that the gyroradius $\rho_{\rm g} = v_\perp/\Omega_{\rm g}$ with gyrofrequency $\Omega_{\rm g} = ZeB/M$ must be small compared to the scale length $L_B \sim R_0$ of the magnetic field ${\bm B}$. Small $\epsilon_\parallel$ means that the distance $2\pi v_\parallel/\Omega_{\rm g}$ traveled during one gyration must be smaller than the parallel wavelength $\lambda_\parallel = 2\pi/k_\parallel$ of the wave ($k_\parallel \sim \hat{\bm b}\cdot\nablab$ with $\hat{\bm b} = {\bm B}/B$). Small $\epsilon_\omega$ means that the wave frequency $\omega$ is far below the cyclotron resonance. When these conditions are satisfied, a charged particle experiences slowly varying fields and, thus, undergoes only slow drifts relative to the magnetic field lines --- with the possible exception of the electric drift ${\bm v}_{\rm E} = {\bm E}\times\hat{\bm b}/B$ that can be allowed to become fast. The purpose of GC theory is to capture those drifts in an efficient and accurate manner. For simplicity, we will use a single ordering parameter
\begin{equation}
\epsilon \sim {\rm max}\{\epsilon_B,\epsilon_\parallel,\epsilon_\omega\}.
\end{equation}

The GC phase space consists of the GC position ${\bm X}_{\rm gc}$, an ignorable gyrophase $\theta$, the GC parallel velocity variable $u \equiv \dot{\bm X}_{\rm gc}\cdot\hat{\bm b} \approx v_\parallel$, and the magnetic moment $\mu$. At lowest order, the latter has the form
\begin{equation}
\mu \equiv \frac{M|{\bm w}|^2}{2 B({\bm X}_{\rm gc},t)},
\end{equation}

\noindent where ${\bm w} \equiv {\bm v}_\perp - {\bm v}_{\rm E}$ is the perpendicular velocity of the particle in the local frame moving with the electric drift velocity ${\bm v}_{\rm E} \equiv {\bm E}\times\hat{\bm b}/B$. The GC phase space Lagrangian $\L(\eta,\dot{\eta};t)$ to order $\O(\epsilon)$ with $\eta = \{{\bm X}_{\rm gc}, u, \mu, \theta)$ is
\begin{align}
\L_{\rm gc} =\;& \left[Ze{\bm A}({\bm X}_{\rm gc},t) + M u \hat{\bm b}({\bm X}_{\rm gc},t)\right]\cdot\dot{\bm X}_{\rm gc} + J\dot\theta - \H_{\rm gc},
\label{eq:lgc}
\end{align}

\noindent with the gyroaction variable $J = \mu B/\Omega_{\rm g}$. The GC Hamiltonian $\H({\bm X}_{\rm gc},u,\mu;t)$ to order $\O(1)$ is
\begin{equation}
\H_{\rm gc} = \frac{M}{2}u^2 + \mu B({\bm X}_{\rm gc},t) + Ze\Phi({\bm X}_{\rm gc},t) - \frac{M}{2}|{\bm v}_{\rm E}({\bm X}_{\rm gc},t)|^2.
\label{eq:hgc}
\end{equation}

\noindent The last two terms in Eq.~(\ref{eq:hgc}) are the finite-Larmor-radius (FLR) expansion of the effective electric potential energy $Ze\Phi({\bm X}_{\rm gc} + \bar{\bm \rho}) + M|{\bm v}_{\rm E}|^2/2$, where
\begin{equation}
\bar{\bm \rho} = \frac{\hat{\bm b}\times{\bm v}_{\rm E}}{\Omega_{\rm g}} = \frac{{\bm E}_\perp}{\Omega_{\rm g} B}
\label{eq:rhobar}
\end{equation}

\noindent is the gyrophase-independent part of the GC position defined as ${\bm x} = {\bm X}_{\rm gc} + {\bm \rho}$ with ${\bm \rho} = \widetilde{\bm \rho}(\theta) + \bar{\bm \rho}$, and where we neglect $Ze\bar{\bm \rho}\cdot\partial_t{\bm A}/ (M v_{\rm E}^2) \sim \O(\omega\alpha/v_{\rm E}) \sim \O(E_\parallel/E_\perp) \sim \O(\epsilon)$. The term $-M|{\bm v}_{\rm E}|^2/2$ is interpreted as a (not-yet-averaged) ponderomotive potential, since it gives rise to a nonlinear force that is proportional to the gradient of an oscillating electric field's intensity. Here, we have $\omega \ll \Omega_{\rm g}$, so the electric field's oscillation emerges in the particle's moving frame of reference as it gyrates through the nonuniform electric field.

%-------------------------------------------------------------------------------
\subsection{Field representation in Boozer coordinates}
\label{apdx:theory_b}

In the present work, the spatial position vector ${\bm X}_{\rm gc}$ in Eq.~(\ref{eq:lgc}) is expressed in the toroidal flux coordinates $(\Psi,\vartheta,\zeta)$ proposed by Boozer \cite{Boozer81, Boozer82}. The attribute `flux coordinates' means that ${\bm B}_\eq\cdot\nablab\Psi_{\rm P} = 0$ in the stationary reference field ${\bm B}_\eq$. Boozer coordinates belong to the class of straight-field-line coordinates (sometimes called `magnetic coordinates') since they satisfy $B_\eq^\zeta/B_\eq^\vartheta = {\rm d}\Psi/{\rm d}\Psi_{\rm P} = q(\Psi_{\rm P})$, so the field helicity (tokamak safety factor) $q$ depends only on $\Psi_{\rm P}$. Although there are reasons to use the toroidal flux function $\Psi$ as a flux label as discussed in Ref.~\onlinecite{Matsuyama13}, that choice becomes arbitrary once one has made the decision to use an {\it unperturbed} flux function. Here, we choose to work with the poloidal flux function $\Psi_{\rm P}$ instead of $\Psi$, and the resulting equations will differ only by factors of $q(\Psi_{\rm P})$.

The short-hand notation $(...)' = {\rm d}(...)/{\rm d}\Psi_{\rm P}$ is used for derivatives of functions that depend on $\Psi_{\rm P}$ only. Partial derivatives of functions that depend on multiple variables are denoted by $(...)'_\eta = \partial_\eta(...)$.

We constrain magnetic fluctuations to the form $\delta{\bm A} = \alpha{\bm B}_\eq$.\footnote{See Eq.~(13) of Ref.~\protect\onlinecite{White84} after which it is stated that perturbations other than $\nablab\times(\alpha{\bm B}_\eq)$ ``contribute only nonresonant distortions of the equilibrium and are unimportant'', citing Rosenbluth {\it et al}.\ {\it Nucl.\ Fusion} {\bf 6} (1966) 297.}
The covariant and contravariant representations of the total field ${\bm B} = {\bm B}_\eq + \delta{\bm B}$ are then\footnote{Here the symbol $\beta_*$ was adopted to represent the covariant component $B_\Psi$ in Boozer coordinates as in the original work \protect\cite{Boozer81}. The same quantity is denoted as $\delta$ by White \& Chance \protect\cite{White84}. Note that a similar symbol $\beta^*$ is used in Appendix B of Ref.~\protect\onlinecite{White84} to represent $B_\Psi$ of the magnetic field expressed in {\it canonical} coordinates.}
\begin{subequations}
\begin{align}
{\bm B} =\;& \underbrace{g\nablab\zeta + I\nablab\vartheta + \beta_*\nablab\Psi}\limits_{\text{reference field}\; {\bm B}_\eq} + \underbrace{\nablab\times(\alpha{\bm B}_\eq)}\limits_{\text{perturbation}}
\label{eq:bflux_co}
\\
{\bm B} =\;& \nablab\times\underbrace{[(\Psi + \alpha I)\nablab\vartheta - (\Psi_{\rm P} - \alpha g)\nablab\zeta + \alpha \beta_*\nablab\Psi]}\limits_{{\bm A} = {\bm A}_\eq + \alpha{\bm B}_\eq},
\label{eq:bflux_contra}
\end{align}
\end{subequations}

\noindent with
\begin{equation}
{\bm B}_\eq = (\partial_\zeta{\bm x} + q^{-1}\partial_\vartheta{\bm x})B^\zeta_\eq.
\end{equation}

\noindent Dotting the co- and contravariant forms of ${\bm B}_\eq$ gives
\begin{equation}
B_\eq^2 \J_{\rm B} = qg + I,
\label{eq:b2j}
\end{equation}

\noindent with the Jacobian for the transformation ${\bm x}\rightarrow(\Psi_{\rm P},\vartheta,\zeta)$
\begin{align}
\J_{\rm B} \equiv \J^{\bm x}_{\Psi_{\rm P}\vartheta\zeta} &= \partial_{\Psi_{\rm P}}{\bm x}\cdot(\partial_\vartheta{\bm x} \times \partial_\zeta{\bm x})  \nonumber
\\
&= \frac{1}{\nablab\Psi_{\rm P}\cdot(\nablab\vartheta\times\nablab\zeta)} = \frac{q}{B^\zeta_\eq}.
\end{align}

\noindent The distinguishing feature of Boozer coordinates is that the covariant field components $g(\Psi_{\rm P})$ and $I(\Psi_{\rm P})$ are independent of $\vartheta$, and so is $B_\eq^2\J_{\rm B}$ in Eq.~(\ref{eq:b2j}).\footnote{Note that for a general equilibrium and general flux coordinates we have $\partial_\zeta I = \partial_\vartheta g$. Consequently, $g$ is independent of $\vartheta$ for an axisymmetric field. See pages 39 and 46 of Ref.~\protect\onlinecite{WhiteTokBook3}.}
Besides simplifying the equations of motion, this property has the consequence that not only the magnetic field lines, but also the diamagnetic lines ${\bm B}_\eq\times\nablab\Psi = (g\partial_\vartheta{\bm x} - I\partial_\zeta{\bm x})q/\J_{\rm B}$ are straight.

The unperturbed plasma current density $\mu_0{\bm J} = \nablab\times{\bm B}_\eq$ has the covariant and contravariant representations
\begin{align}
\mu_0{\bm J} &= (g' - q\beta_{*\zeta}')\nablab\zeta\times\nablab\Psi_{\rm P} + (I' - q\beta_{*\vartheta}')\nablab\Psi_{\rm P}\times\nablab\vartheta \nonumber
\\
&= \frac{-(g' - q\beta_{*\zeta}')\partial_\vartheta{\bm x} + (I' - q\beta_{*\vartheta}')\partial_\zeta{\bm x}}{\J_{\rm B}}.
\end{align}

\noindent Its parallel component $J_\parallel \equiv {\bm J}\cdot{\bm B}_\eq/B_\eq$ has the form
\begin{equation}
\frac{\mu_0 J_\parallel}{B_\eq} = \frac{gI' - Ig' - gq\beta_{*\vartheta}' + Iq\beta_{*\zeta}'}{qg + I}.
\label{eq:jpar}
\end{equation}

\noindent The terms $gI' - Ig'$ can be shown to represent the force-free current and the terms containing $\beta_*$ measure the effect of the plasma pressure gradient, known as Pfirsch-Schl\"{u}ter current \cite{Boozer81} (p.~40 of Ref.~\onlinecite{WhiteTokBook3}). Geometrically, $\beta_*$ is manifested in the coordinates' nonorthogonality. In the present work, we consider an axisymmetric reference field ${\bm B}_\eq$, where $\beta_*(\Psi_{\rm P},\vartheta)$ has the form
\begin{subequations}
\begin{align}
\beta_* =\;& -I(\Psi_{\rm P}) \frac{\nablab\Psi\cdot\nablab\vartheta}{|\nablab\Psi|^2} = -\frac{I(\Psi_{\rm P})}{q(\Psi_{\rm P})} \frac{\nablab\Psi_{\rm P}\cdot\nablab\vartheta}{|\nablab\Psi_{\rm P}|^2}
\label{eq:delta1}
\\
=\;& \frac{\partial_\Psi{\bm x}\cdot\partial_\vartheta{\bm x}}{\J_{\rm B}} = \frac{g(\Psi_{\rm P})}{q(\Psi_{\rm P})}|\nablab\zeta|^2 \partial_\Psi{\bm x}\cdot\partial_\vartheta{\bm x}.
\label{eq:delta2}
\end{align}
\end{subequations}

\noindent The form in Eq.~(\ref{eq:delta1}) follows from ${\bm B}_\eq\cdot\nablab\Psi = 0$. The form in Eq.~(\ref{eq:delta2}) follows from $\beta_* = {\bm B}_\eq\cdot\partial_\Psi{\bm x}$ with the contravariant representation (\ref{eq:bflux_contra}) and $B_\eq^\zeta = g|\nablab\zeta|^2$.

Note that $\nablab\cdot{\bm B}_\eq = \nablab\beta_*\cdot\nablab\Psi = 0$ implies that
\begin{equation}
-\frac{\nablab\Psi_{\rm P}\cdot\nablab\vartheta}{|\nablab\Psi_{\rm P}|^2} = \frac{q\beta_*}{I} = \frac{q\beta_{\Psi_{\rm P}}'}{I' + q\beta_{*\vartheta}'}.
\end{equation}

\noindent Numerical inaccuracies that break this exact relationship between $\beta_*$, $\beta_{\Psi_{\rm P}}'$ and  $\beta_{*\vartheta}'$ would consequently violate the solenoidal condition $\nablab\cdot{\bm B}_\eq = 0$. The usual choice to ignore $\beta_*$ in GC models has the benefit of avoiding this risk.

%-------------------------------------------------------------------------------
\subsection{Normalization}
\label{apdx:theory_norm}

\vspace{-0.2cm}
In the following, the magnetic field and time are normalized as
\begin{equation}
\hat{B} = B/B_0, \quad \hat{t} = t\Omega_{\rm g0}, \quad \text{with} \;\; \Omega_{\rm g0} = ZeB_0/M,
\end{equation}

\noindent so that $\hat{\Omega} = \hat{B}$. Velocities, potentials and energies are then normalized as
\begin{equation}
\hat{v} = \frac{v}{\Omega_{\rm g0}}, \quad \hat{\bm A} = \frac{\bm A}{B_0}, \quad \hat{\Phi} = \frac{\Phi}{B_0 \Omega_{\rm g0}}, \quad \hat{\H} = \frac{\H}{M\Omega_{\rm g0}^2}.
\end{equation}

\noindent In the {\tt ORBIT} code \cite{White84}, a certain reference cyclotron frequency $\Omega_{\rm g0,ref} = Q_{\rm ref}e B_0/M_{\rm ref}$ is used for normalization; usually deuteron. Equations of motion for other species can then be obtained via the following substitutions:
\begin{subequations}
\begin{align}
&\hat{t} \rightarrow \hat{t}\frac{\Omega_{\rm g0}}{\Omega_{\rm g,ref}} = \hat{t}\frac{Z_{\rm Q}}{A_{\rm M}},
\\
&\left(\frac{1}{\hat{t}},\rho,\hat{v},\hat{\Phi}\right) \rightarrow \left(\frac{1}{\hat{t}},\rho,\hat{v},\hat{\Phi}\right)\frac{A_{\rm M}}{Z_{\rm Q}},
\\
& (\hat{\H}_{\rm gc},\hat{\L}_{\rm gc},\hat{\mu}) \rightarrow (\hat{\H}_{\rm gc},\hat{\L}_{\rm gc},\hat{\mu})\frac{A_{\rm M}}{Z_{\rm Q}^2},
\end{align}
\end{subequations}

\noindent with $Z_{\rm Q} \equiv Q/Q_{\rm ref}$ and $A_{\rm M} \equiv M/M_{\rm ref}$.

The normalized GC phase-space Lagrangian and Hamiltonian are
\begin{subequations}
\begin{align}
&\underbrace{\frac{\L_{\rm gc}}{M\Omega_{\rm g0}^2}}\limits_{\hat{\L}} = \underbrace{\frac{{\bm A}}{B_0}\cdot\frac{\dot{\bm X}_{\rm gc}}{\Omega_{\rm g0}}}\limits_{\hat{\bm A}\cdot\dot{\bm X}_{\rm gc}} + \underbrace{\frac{u}{\Omega_{\rm g}}\frac{{\bm B}}{B_0}\cdot\frac{\dot{\bm X}_{\rm gc}}{\Omega_{\rm g0}}}\limits_{\rho_\parallel \hat{\bm B}\cdot\hat{\dot{\bm X}}_{\rm gc} = \rho_\parallel^2 \hat{B}^2} + \underbrace{\frac{\mu B_0}{M\Omega_{\rm g0}^2}\frac{\dot\theta}{\Omega_{\rm g0}}}\limits_{\hat{\mu}\hat{\dot\theta}} - \underbrace{\frac{\H_{\rm gc}}{M\Omega_{\rm g0}^2}}\limits_{\hat{\H}},
\label{eq:lgc_nrm}
\\
&\underbrace{\frac{\H_{\rm gc}}{M\Omega_{\rm g0}^2}}\limits_{\hat{\H}} = \frac{1}{2}\underbrace{\left(\frac{u}{\Omega_{\rm g}}\frac{B}{B_0}\right)^2}\limits_{\rho_\parallel^2 \hat{B}^2} + \underbrace{\frac{\mu B_0}{M\Omega_{\rm g0}^2} \frac{B}{B_0}}\limits_{\hat{\mu} \hat{B}} + \underbrace{\frac{\Phi}{B_0\Omega_{\rm g0}}}\limits_{\hat{\Phi}} - \frac{1}{2}\underbrace{\left|\frac{{\bm v}_{\rm E}}{\Omega_{\rm g0}}\right|^2}\limits_{|\hat{\bm v}_{\rm E}|^2};
\label{eq:hgc_nrm}
\end{align}
\end{subequations}

\noindent where $\rho_\parallel = \dot{\bm X}_{\rm gc}\cdot{\bm B}/(B\Omega_{\rm g}) = u/\Omega_{\rm g} = \hat{u}/\hat{B}$. Omitting the hats and the subscripts `gc', we have
\begin{subequations}
\begin{align}
\L =\;& ({\bm A} + \rho_\parallel {\bm B})\cdot\dot{\bm X} + \mu\dot\theta - \H,
\label{eq:lgcn}
\\
\H =\;& \rho_\parallel^2 B^2/2 + \mu B + \Phi - |{\bm v}_{\rm E}|^2/2.
\label{eq:hgcn}
\end{align}
\end{subequations}
%\vspace{0.1cm}

%-------------------------------------------------------------------------------
\subsection{Phase-space Lagrangian formulation of GC equations in the long-wavelength limit}
\label{apdx:theory_gce}

We assume long-wavelength perturbations, with gradient scale length $L_\alpha$ much larger than the gyroradius, but shorter than or comparable to the scale length $L_B \sim R_0$ of the reference field:
\begin{equation}
|\nablab\alpha| \sim \alpha/L_\alpha, \quad \text{with} \quad L_B \gtrsim L_\alpha \gg \rho_\parallel.
\end{equation}%\vspace{0.1cm}

\noindent Using $\nablab\Psi = q(\Psi_{\rm P})\nablab\Psi_{\rm P}$ and $\dot{\bm X} = \dot\zeta\partial_\zeta{\bm x} + \dot\vartheta\partial_\vartheta{\bm x} + \dot\Psi_{\rm P}\partial_{\Psi_{\rm P}}{\bm x}$, substitution into (\ref{eq:lgcn}) yields the Lagrangian
\begin{subequations}
\begin{align}
\L =\;& (\rho_\parallel + \alpha){\bm B}_\eq\cdot\dot{\bm X} + {\bm A}_\eq\cdot\dot{\bm X}
\label{eq:lgc_tor1}
\\
& + \mu\dot\theta - \H + {\cgray \rho_\parallel[\nablab\times(\alpha{\bm B}_\eq)]\cdot\dot{\bm X}} \nonumber
\\
\approx\;& [(\rho_\parallel + \alpha) g - \Psi_{\rm P}]\dot\zeta + [(\rho_\parallel + \alpha) I + \Psi]\dot\vartheta
\label{eq:lgc_tor}
\\
& + [(\rho_\parallel + {\cred \alpha}) q \beta_*] \dot\Psi_{\rm P} + \mu\dot\theta - \H \nonumber
\end{align}
\end{subequations}

\noindent where we have omitted terms of order $\O(\rho_\parallel^2 \alpha/L_B)$, $\O(\rho_\parallel^2 \alpha/L_\alpha \times U_\perp/u)$ and higher, namely the last term in Eq.~(\ref{eq:lgc_tor1}) printed gray. This also means that
\begin{equation}
\rho_\parallel = \frac{u}{B_\eq}\left[1 + \O\left(\frac{\alpha}{L_B}, \frac{\alpha}{L_\alpha}\frac{\dot{X}_\perp}{u}\right)\right] \approx \frac{u}{B_\eq}.
\label{eq:rhopar}
\end{equation}

\noindent Similarly, Eq.~(\ref{eq:hgcn}) for the Hamiltonian becomes
\begin{equation}
\H = \rho_\parallel^2 B_\eq^2/2 + \mu B_\eq + \Phi - |{\bm v}_{\rm E}|^2/2.
\end{equation}

The Euler-Lagrange equations ${\rm d}_t\partial_{\dot{\eta}} \L = \partial_\eta \L,$ for the GC phase-space Lagrangian $\L({\bm \eta},\dot{\bm\eta}; t)$ in Eq.~(\ref{eq:lgc_tor}) with coordinates ${\bm \eta} = \{\Psi_{\rm P},\vartheta,\zeta,\rho_\parallel,\mu,\theta\}$ are then
\begin{widetext}
\begin{equation}%\hspace{-0.8cm}
\begin{array}{llcl}
\eta & {\rm d}_t\partial_{\dot{\eta}} \L &=& \partial_\eta \L,
\\
\rho_\parallel:& 0 &=& q \beta_* \dot\Psi_{\rm P} + I\dot\vartheta + g\dot\zeta - \H'_{\rho_\parallel}, \\
\Psi_{\rm P}:& q \beta_* \dot\rho_\parallel + q\beta_*\partial_t\alpha &=& - \H'_{\Psi_{\rm P}} \\
& {\cgray + \rho_\alpha(q'_{\Psi_{\rm P}}\beta_* + q\beta'_{*\Psi_{\rm P}})\dot\Psi_{\rm P}}  {\cgray + \alpha'_{\Psi_{\rm P}} q \beta_*\dot\Psi_{\rm P}} && {\cgray +\rho_\alpha(q'_{\Psi_{\rm P}}\beta_* + q\beta'_{*\Psi_{\rm P}})\dot\Psi_{\rm P}}  {\cgray + \alpha'_{\Psi_{\rm P}} q \beta_*\dot\Psi_{\rm P}} \\
& + \rho_\alpha q \beta'_{*\vartheta}\dot\vartheta {\cred + \alpha'_\vartheta q\beta_*\dot\vartheta}  {\cred + \alpha'_\zeta q \beta_*\dot\zeta} && + (\rho_\alpha I'_{\Psi_{\rm P}} + I\alpha'_{\Psi_{\rm P}} + q)\dot\vartheta + (\rho_\alpha g'_{\Psi_{\rm P}} + g\alpha'_{\Psi_{\rm P}} - 1)\dot\zeta, \\
\vartheta:& \dot\rho_\parallel I + I\partial_t\alpha &=& - \H'_\vartheta \\
& + (\rho_\alpha I'_{\Psi_{\rm P}} + I\alpha'_{\Psi_{\rm P}} + q)\dot\Psi_{\rm P} && \rho_\alpha q\beta'_{*\vartheta}\dot\Psi_{\rm P} {\cred + \alpha'_\vartheta q\beta_* \dot\Psi_{\rm P}} \\
& {\cgray + \alpha'_\vartheta I \dot\vartheta} + \alpha'_\zeta I \dot\zeta && {\cgray + \alpha'_\vartheta I \dot\vartheta} + \alpha'_\vartheta g \dot\zeta, \\
\zeta:& \dot\rho_\parallel g + g\partial_t\alpha &=& - \H'_\zeta \\
& + (\rho_\alpha g'_{\Psi_{\rm P}} + g\alpha'_{\Psi_{\rm P}} - 1)\dot\Psi_{\rm P} && {\cred + \alpha'_\zeta q\beta_* \dot\Psi_{\rm P}} \\
& + g\alpha'_\vartheta\dot\vartheta {\cgray + g\alpha'_\zeta\dot\zeta} && + I\alpha'_\zeta\dot\vartheta {\cgray + g\alpha'_\zeta\dot\zeta}, \\
\theta:& \dot\mu &=& 0;
\end{array}
\label{eq:variation}
\end{equation}
\end{widetext}

\noindent where $(...)'_\eta \equiv \partial_\eta(...)$ as before, and where we wrote
\begin{equation}
\rho_\alpha \equiv \rho_\parallel + \alpha
\end{equation}

\noindent for compact notation. Terms that cancel are printed gray. Written in matrix form, Eq.~(\ref{eq:variation}) becomes
\arraycolsep=2.0pt\def\arraystretch{1.0}
\begin{equation}
\left[\begin{array}{cccc}
0 & q\beta_* & I & g \\
-q\beta_* & 0 & F & C \\
-I & -F & 0 & K \\
-g & -C & -K & 0
\end{array}\right]
\left[\begin{array}{cccc}
\dot\rho_\parallel \\ \dot\Psi_{\rm P} \\ \dot\vartheta \\ \dot\zeta
\end{array}\right]
 =
\left[\begin{array}{cccc}
\H'_{\rho_\parallel} \\
\H'_{\Psi_{\rm P}} + q\beta_*\partial_t\alpha \\
\H'_\vartheta + I\partial_t\alpha \\
\H'_\zeta + g\partial_t\alpha
\end{array}\right],
\label{eq:variation_mat}
\end{equation}

\noindent with
\begin{subequations}
\begin{align}
F \equiv& q + \rho_\alpha (I'_{\Psi_{\rm P}} - q \beta'_{*\vartheta}) + I\alpha'_{\Psi_{\rm P}} {\cred - \alpha'_\vartheta q \beta_*},
\label{eq:fck_f}
\\
C \equiv& -1 + \rho_\alpha g'_{\Psi_{\rm P}} + g\alpha'_{\Psi_{\rm P}} {\cred - \alpha'_\zeta q\beta_*},
\label{eq:fck_c}
\\
K \equiv& g\alpha'_\vartheta - I\alpha'_\zeta.
\end{align}
\label{eq:fck}
\end{subequations}

\noindent Inversion of Eq.~(\ref{eq:variation_mat}) gives
\arraycolsep=2.0pt\def\arraystretch{1.0}
\begin{equation}
\left[\begin{array}{cccc}
\dot\rho_\parallel \\ \dot\Psi_{\rm P} \\ \dot\vartheta \\ \dot\zeta
\end{array}\right]
 =
\frac{1}{D} \left[\begin{array}{cccc}
0 & -K & C & -F \\
K & 0 & -g & I \\
-C & g & 0 & -q\beta_* \\
F & -I & q\beta_* & 0
\end{array}\right]
\left[\begin{array}{cccc}
\H'_{\rho_\parallel} \\
\H'_{\Psi_{\rm P}} + q\beta_*\partial_t\alpha \\
\H'_\vartheta + I\partial_t\alpha \\
\H'_\zeta + g\partial_t\alpha
\end{array}\right],
\label{eq:eom_matrix}
\end{equation}

\noindent with
\begin{equation}
D \equiv gF - IC + K q\beta_* = gq + I + \rho_\alpha(g I'_{\Psi_{\rm P}} - I g'_{\Psi_{\rm P}} - g q \beta'_{*\vartheta}).
\label{eq:d}
\end{equation}

\noindent Note that all terms containing derivatives of $\alpha$ cancel inside the denominator $D$. In the following equations (\ref{eq:dH}) and (\ref{eq:eom_lgc_booz}), we let $B_\eq \rightarrow B$, omitting the subscript `$\eq$' since field perturbations appear explicitly as $\alpha$ and $\Phi$. The derivatives of the Hamiltonian $\H$ are then
\begin{subequations}
\begin{align}
\H'_{\rho_\parallel} =\;& \rho_\parallel B^2
\\
\H'_{\Psi_{\rm P}} =\;& (\rho_\parallel^2 B + \mu) B'_{\Psi_{\rm P}} + \Phi'_{\Psi_{\rm P}} - \partial_{\Psi_{\rm P}}|{\bm v}_{\rm E}|^2/2
\\
\H'_\vartheta =\;& (\rho_\parallel^2 B + \mu) B'_\vartheta + \Phi'_\vartheta - \partial_\vartheta|{\bm v}_{\rm E}|^2/2
\\
\H'_\zeta =\;& \underbrace{\cgray (\rho_\parallel^2 B + \mu) B'_{{\rm rip},\zeta}}\limits_{\text{with toroidal field ripple}} + \Phi'_\zeta - \partial_\zeta|{\bm v}_{\rm E}|^2/2,
\end{align}
\label{eq:dH}
\end{subequations}

\noindent with {\tt ORBIT} containing an option to include toroidal field ripple $B_{\rm rip}$. The resulting equations of motion are
\begin{subequations}
\begin{align}
D\dot\rho_\parallel =\;& (-K\partial_{\Psi_{\rm P}} + C\partial_\vartheta - F\partial_\zeta)\H - D\partial_t\alpha,
\label{eq:eom_lgc_booz_rho}
\\
D\dot\Psi_{\rm P} =\;& K\rho_\parallel B^2 + (-g\partial_\vartheta + I\partial_\zeta)\H,
\\
D\dot\vartheta =\;& -C\rho_\parallel B^2 + (g\partial_{\Psi_{\rm P}} - q\beta_*\partial_\zeta)\H,
\\
D\dot\zeta =\;& F\rho_\parallel B^2 + (-I\partial_{\Psi_{\rm P}} + q\beta_*\partial_\vartheta)\H.
\end{align}\vspace{-0.2cm}
\label{eq:eom_lgc_booz}
\end{subequations}

\noindent These equations are obtained by combining Eqs.~(3.19)--(3.22) on p.~78 of Ref.~\onlinecite{WhiteTokBook3} and Eqs.~(3.93)--(3.96) on p.~106f of Ref.~\onlinecite{WhiteTokBook3}, with the addition of the ponderomotive term $|{\bm v}_{\rm E}|^2/2$ in $\H$ from Ref.~\onlinecite{Cary09} (absent from {\tt ORBIT}).

%-------------------------------------------------------------------------------
\subsection{Conservation of GC phase space volume}
\label{apdx:theory_liouville}

It can be readily verified that the quantity $D$ in Eq.~(\ref{eq:d}) satisfies the divergence equation
\begin{equation}
\dot{D} = \partial_t D + \nablab\cdot(D\dot{\bm X}) + \partial_{\rho_\parallel}(D\dot{\rho}_\parallel) = 0.
\label{eq:liouville}
\end{equation}

\noindent The individual terms are
\begin{subequations}
\begin{align}
\partial_t D =& {\cgray \partial_t\alpha(gI' - Ig' - gq\beta_{*\vartheta}')},
\\
\partial_{\rho_\parallel}(D\dot\rho_\parallel) =& {\cgray -\partial_t\alpha (gI' - Ig' - gq\beta_{*\vartheta}')} \\
& {\cgray - K\H''_{\rho_\parallel\Psi_{\rm P}} + C\H''_{\rho_\parallel\vartheta} - F\H''_{\rho_\parallel\zeta}} \nonumber \\
& {\cg g'_{\Psi_{\rm P}}\H'_\vartheta  - (I' - q\beta_{*\vartheta}')\H'_\zeta},
\\
\partial_{\Psi_{\rm P}}(D\dot{\Psi}_{\rm P}) =& \partial_{\Psi_{\rm P}}\left( K\H'_{\rho_\parallel} - g\H'_\vartheta + I\H'_\zeta\right) \nonumber \\
\rightarrow& K'_{\Psi_{\rm P}} \H'_{\rho_\parallel} {\cg\, -\, g'\H'_\vartheta + I'\H'_\zeta},
\\
\partial_\vartheta(D\dot{\vartheta}) =& \partial_\vartheta\left( -C\H'_{\rho_\parallel} + g\H'_{\Psi_{\rm P}} - q\beta_*\H'_\zeta\right) \nonumber \\
\rightarrow& -C'_\vartheta \H'_{\rho_\parallel} {\cg\, -\, q\beta_{*\vartheta}'\H'_\zeta},
\\
\partial_\zeta(D\dot{\zeta}) =& \partial_\zeta\left( F\H'_{\rho_\parallel} - I\H'_{\Psi_{\rm P}} + q\beta_*\H'_\vartheta\right) \nonumber \\
\rightarrow& F_\zeta' \H'_{\rho_\parallel},
\end{align}
\end{subequations}

\noindent where we have made use of the immediately obvious fact that all the second-order derivatives of $\H$ and the two terms containing $\partial_t\alpha$ cancel, so we have printed them gray or omitted them after the arrows. One can also see that the green terms containing $\H'_{\Psi_{\rm P}}$, $\H'_\vartheta$ and $\H'_\zeta$ cancel pairwise. What remains are the first terms of the last three equations, which contain $\H'_{\rho_\parallel}$. After expanding the factors multiplying $\H'_{\rho_\parallel}$,
\begin{subequations}
\begin{align}
K'_{\Psi_{\rm P}} =& g'\alpha'_\vartheta - I'\alpha'_\zeta + g\alpha''_{\Psi_{\rm P}\vartheta} - I\alpha''_{\Psi_{\rm P}\zeta}, \\
- C'_\vartheta =& -g'\alpha_\vartheta - g\alpha''_{\Psi_{\rm P}\vartheta} + \alpha''_{\vartheta\zeta} q\beta_* + \alpha'_\zeta q \beta_{*\vartheta}', \\
F'_\zeta =& \alpha'_\zeta(I' - q\beta'_{*\vartheta}) + I\alpha''_{\Psi_{\rm P}\zeta} - \alpha''_{\vartheta\zeta}q \beta_*.
\end{align}
\end{subequations}

\noindent it is clear that the remaining terms cancel as well, which proves that $\dot{D} = 0$.

If $D$ in Eq.~(\ref{eq:d}) was the Jacobian of a (time-dependent!) phase space coordinate transformation, then this coordinate transformation in combination with the GC equations of motion (\ref{eq:eom_lgc_booz}) satisfies the Liouville theorem, meaning that the associated Hamiltonian flow conserves phase space volume. Let us now compare $D$ with the actual Jacobian for our noncanonical GC model in Boozer coordinates.

The GC coordinates ${\bm \eta} = \{{\bm X},u,\mu,\theta\}$ appearing in Eq.~(\ref{eq:lgc}) contain noncanonical velocity variables whose Jacobian factor $B^*_\parallel = \hat{\bm b}\cdot{\bm B}^*$ (first derived by Littlejohn \cite{Littlejohn83}) satisfies the phase space conservation law (\ref{eq:liouville}) (cf., p.~704 of Ref.~\onlinecite{Cary09}). Here, we use $\rho_\parallel = u/\Omega_{\rm g} = \hat{u}/\hat{B}$ instead of $u$ as a parallel velocity coordinate, so our coordinate transformation has the form (with normalizations shown explicitly by hat symbols)
\begin{equation}
{\rm d}^3{\bm x}\frac{{\rm d}^3{\bm v}}{\Omega_{\rm g0}^3} = {\rm d}^3{\bm X}\frac{{\rm d}\mu B_0}{M\Omega_{\rm g0}^2}\hat{B}{\rm d}\rho_\parallel \frac{B^*_\parallel}{B_0} = {\rm d}^3{\bm X} {\rm d}\hat{\mu}{\rm d}\rho_\parallel \hat{B}\hat{\bm B}_\parallel,
\end{equation}

\noindent with velocity space Jacobian
\begin{equation}
\hat{\J}^{\bm v}_{\rho_\parallel\mu\theta} = \hat{B}\hat{B}^*_\parallel.
\end{equation}

\noindent In the following, we will again omit the normalization hats. The auxiliary field ${\bm B}^*$ has the form
\begin{equation}
{\bm B}^* = {\bm B} + \rho_\parallel (\nablab\times{\bm B} + \hat{\bm b}\times\nablab B).
\end{equation}

\noindent Substituting ${\bm B} = {\bm B}_\eq + \nablab\times\alpha{\bm B}_\eq$ and ignoring small terms as we did earlier in $\L$ and $\H$, we obtain
\begin{align}
{\bm B}^* \approx&\; {\bm B}_\eq + (\rho_\parallel + \alpha)\nablab\times{\bm B}_\eq + \hat{\bm b}\times\nablab B - {\bm B}_\eq\times\nablab\alpha \nonumber
\\
& + \O(\alpha^2,\rho_\parallel\alpha/L_\alpha,\rho_\parallel\alpha/L_B)
\end{align}

\noindent so that
\begin{equation}
B^*_\parallel \approx B_\eq + \rho_\alpha\mu_0 J_\parallel/B_\eq.
\label{eq:bstar_approx}
\end{equation}

\noindent with $\rho_\alpha = \rho_\parallel + \alpha$ and $\mu_0\J_\parallel = \hat{\bm b}_\eq\cdot(\nablab\times{\bm B}_\eq)$ as before.

The Boozer coordinates that we use for the spatial positions are also noncanonical. Their Jacobian is
\begin{equation}
\J_{\rm B} \equiv \J^{\bm x}_{\Psi_{\rm P}\vartheta\zeta}(\Psi_{\rm P},\vartheta) = (gq + I)/B^2_\eq.
\end{equation}

\noindent The combined Jacobian $\J$ for our noncanonical set of coordinates ${\bm \eta} = \{\Psi_{\rm P},\vartheta,\zeta,\rho_\parallel,\mu,\theta\}$ is then
\begin{align}
\J &= \J^{\bm v}_{\rho_\parallel\mu\theta} \J^{\bm x}_{\Psi_{\rm P}\vartheta\zeta} = (gq + I)B^*_\parallel/B_\eq \nonumber
\\
&\approx gq + I + \rho_\parallel(gI' - Ig' - gq\beta_{*\vartheta}')
\label{eq:j_approx}
\end{align}

\noindent where we have used Eq.~(\ref{eq:jpar}) for the parallel current $\mu_0 J_\parallel/B$ appearing inside $B^*_\parallel$ given by Eq.~(\ref{eq:bstar_approx}). It is now clear that our approximate Jacobian in Eq.~(\ref{eq:j_approx}) is identical to $D$ from Eq.~(\ref{eq:d})
\begin{equation}
\J = D + \O(\alpha^2,\rho_\parallel\alpha/L_\alpha,\rho_\parallel\alpha/L_B).
\end{equation}

Two points in the above derivation are of importance for the topic of the present work:
\begin{itemize}
\item  We were able to demonstrate phase space conservation while retaining $\beta_*$.

\item  The derivation of phase space conservation for the present GC model relied on the mutual cancellation of the small terms containing $\alpha'_\vartheta\beta_*$ and $\alpha'_\zeta\beta_*$ that are printed red in Eq.~(\ref{eq:fck}). Phase space conservation is ensured as long as both terms are retained or neglected together. It is broken, however, when only one term is omitted, and the consequences were shown in Figs.~\ref{fig:12}--\ref{fig:15} of Section~\ref{sec:results_break}.
\end{itemize}

%-------------------------------------------------------------------------------
\subsection{Conservation of rotating frame energy $\E'$}
\label{apdx:theory_erot}

Let us now examine the relation between changes in energy $\H$ and momentum $\P_\zeta = (\rho_\parallel + \alpha)g - \Psi_{\rm P}$. Their time derivatives are
\begin{align}
\dot\H =\;& \H'_{\rho_\parallel}\dot\rho_\parallel + \H'_{\Psi_{\rm P}}\dot\Psi_{\rm P} +\H'_\vartheta\dot\vartheta + \H'_\zeta\dot\zeta + \partial_t\H,
\label{eq:dh_dt}
\\
\dot\P_\zeta =\;& g\dot\rho_\parallel + g\alpha'_\vartheta\dot\vartheta + g\alpha'_\zeta\dot\zeta + g\partial_t\alpha \nonumber
\\
&+ \underbrace{[(\rho_\parallel + \alpha)g'_{\Psi_{\rm P}} - 1 + g\alpha'_{\Psi_{\rm P}}]}\limits_{C + \alpha'_\zeta q \beta_*}\dot\Psi_{\rm P}.
\end{align}

\noindent Note that, in the absence of any explicit time dependence, the antisymmetry of the matrix giving the time derivatives in Eq.~(\ref{eq:eom_matrix}) guarantees that energy is conserved, because all terms in the total time derivative of $\H$ in Eq.~(\ref{eq:dh_dt}) cancel \cite{White13c}.

Substituting the equations of motion (\ref{eq:eom_lgc_booz}) one obtains, after some algebra and many cancellations,
\begin{subequations}
\begin{align}
\dot\H =\;& \partial_t\H - \H'_{\rho_\parallel}\partial_t\alpha,
\label{eq:dh_dt_expanded}
\\
\dot\P_\zeta =\;& -\H'_\zeta + \H'_{\rho_\parallel}\alpha'_\zeta.
\label{eq:dpz_dt_expanded}
\end{align}
\label{eq:dh-dpz_dt_expanded}
\end{subequations}%\vspace{-0.3cm}

\noindent Note that the accuracy of the simulation can be enhanced by constraining the dynamics with Eqs.~(\ref{eq:dh_dt_expanded}) and (\ref{eq:dpz_dt_expanded}) for $\dot\H$ and $\dot{P}_\zeta$ as done in Ref.~\cite{White13c}. However, the price one pays in computation time rarely justifies the gain in accuracy.

Now consider the case where $\Phi$ and $\alpha$ (with a $\pi/2$ phase shift) depend on the toroidal angle $\zeta$ and time $t$ as
\begin{align}
\Phi &= \sum_{k=0,1,2...}\frac{1}{2}\widetilde{\Phi}_k(R,z)e^{in_k\zeta - i\omega_k t} + {\rm c.c.}, \nonumber
\\
&= \sum_{k=0,1,2...}\frac{1}{2}\widetilde{\Phi}_k(R,z)e^{(in_0\zeta - i\omega_0 t)\kappa_k} + {\rm c.c.},
\end{align}

\noindent so the mode has a time-independent amplitude and poloidal profile $\widetilde{\Phi}(R,z)$, a constant fundamental frequency $\omega_0$, and a single fundamental toroidal mode number $n_0$. The tilde indicates that the poloidal mode structure is a complex function, here of the form $\widetilde{\Phi}(\psi_{\rm P}(R,z),\vartheta(R,z)) = \Phi_0\sum_m\hat\Phi_m(\psi_{\rm P})e^{i\Theta_{0,m} - im\vartheta}$. Note that the waveform may be arbitrarily distorted by superimposing harmonics $k = 1,2,...$ satisfying $\omega_k/n_k = \omega_0/n_0$ for all $k$, so that $\omega_k = \kappa_k\omega_0$ and $n_k = \kappa_k n_0$ with $\kappa_0 = 1$ and arbitrary integer $\kappa_{k>0}$. The ponderomotive potential has then the form
\begin{align}
|{\bm v}_{\rm E}|^2 = \frac{1}{4}\sum_k\sum_{k'}&\left(\widetilde{\bm v}_{{\rm E}k}\cdot\widetilde{\bm v}_{{\rm E}k'} e^{(\kappa_k + \kappa_{k'})(in_0\zeta - i\omega_0 t)}\right.
\\
& + \left.\widetilde{\bm v}_{{\rm E}k}\cdot\widetilde{\bm v}_{{\rm E}k'}^\dagger e^{(\kappa_k - \kappa_{k'})(in_0\zeta - i\omega_0 t)}\right) + {\rm c.c.}, \nonumber
\end{align}

\noindent where the dagger indicates a complex conjugate. Evidently,
\begin{equation}
n_0\partial_t|{\bm v}_{\rm E}|^2 = -\omega_0\partial_\zeta|{\bm v}_{\rm E}|^2.
\label{eq:dpond}
\end{equation}

\noindent Equations~(\ref{eq:dh_dt_expanded}) and (\ref{eq:dpz_dt_expanded}) then become
\begin{subequations}
\begin{align}
\dot\H =\;& -i\omega_0\left(\sum_k\kappa_k(\H + |{\bm v}_{\rm E}|^2/2)_k\right) + \partial_t|{\bm v}_{\rm E}|^2/2 \nonumber
\\
& + i\omega_0\H'_{\rho_\parallel}\sum_k\kappa_k\alpha_k,
\label{eq:dh_dt_mode}
\\
\dot\P_\zeta =\;& -in_0\left(\sum_k\kappa_k(\H + |{\bm v}_{\rm E}|^2/2)_k\right) - \partial_\zeta|{\bm v}_{\rm E}|^2/2 \nonumber
\\
&+ in_0\H'_{\rho_\parallel}\sum_k\kappa_k\alpha_k.
\label{eq:dpz_dt_mode}
\end{align}
\label{eq:dh_dpz_dt_mode}
\end{subequations}

\noindent Multiplying these two equations by $n_0$ and $\omega_0$, respectively, and using Eq.~(\ref{eq:dpond}), one finds that the energy $\E'$ in the wave's rotating frame of reference $\zeta' = \zeta - \omega_0 t/n_0$ is conserved \cite{Hsu94}:
\begin{equation}
\omega_0\dot\P_\zeta = n_0\dot\H \quad \Rightarrow \quad \E' = \E - \frac{\omega_0}{n_0}\P_\zeta = {\rm const}.,
\label{eq:erot}
\end{equation}

\noindent with $\E = \H$.

Two points in the above derivation are of importance for the topic of the present work:
\begin{itemize}
\item  $\beta_*$ does not appear explicitly in Eq.~(\ref{eq:dh-dpz_dt_expanded}), so that retention of $\beta_*$ in the GC phase-space Lagrangian does not upset the conservation of the rotating frame energy $\E'$ in Eq.~(\ref{eq:erot}). This is consistent with our simulation results reported in Section~\ref{sec:results_bench}, where we showed that the unabridged GC Lagrangian including $\beta_*$ yields dynamics with a phase space topology very similar to what one obtains with the GC Lagrangian in canonical form.

\item  The cancellations leading to Eq.~(\ref{eq:dpz_dt_expanded}) rely on the presence of the small terms containing $\alpha'_\vartheta\beta_*$ and $\alpha'_\zeta\beta_*$ that are printed red in Eqs.~(\ref{eq:fck_f}) and (\ref{eq:fck_c}). This explains the results in Figs.~\ref{fig:08}--\ref{fig:11} of Section~\ref{sec:results_break}. Meanwhile, phase space conservation is satisfied even without these terms as demonstrated in Appendix~\ref{apdx:theory_liouville} above.
\end{itemize}

%-------------------------------------------------------------------------------
\subsection{Hamiltonian formulation}
\label{apdx:theory_h}

The full orbit dynamics described by the Newton-Lorentz equation with prescribed (static or fluctuating) electromagnetic fields constitute a Hamiltonian system. Hamiltonian systems are guaranteed to satisfy phase space conservation. The derivation of GC theory makes use of approximations that bear the danger of breaking the Hamiltonian character of the system. It is hence desirable to construct GC model in a perfectly conservative fashion. This can be guaranteed when the Lagrangian has the canonical form
\begin{equation}
\L_{\rm c}({\bm \Theta},{\bm P},\dot{\bm \Theta},\dot{\bm P},t) = P_k\dot{\Theta}^k - H({\bm P},{\bm \Theta},t).
\end{equation}

\noindent where ${\bm \Theta}$ represents three `angle' coordinates and ${\bm P}$ the corresponding `actions' (canonical momenta).

The Lagrangian in Eq.~(\ref{eq:lgc_tor}),
\begin{align}
\L =\;& \underbrace{[(\rho_\parallel + \alpha) g - \Psi_{\rm P}]}\limits_{\P_\zeta}\dot\zeta + \underbrace{[(\rho_\parallel + \alpha) I + \Psi]}\limits_{\P_\vartheta}\dot\vartheta \nonumber
\\
& + [(\rho_\parallel + \alpha) q \beta_*] \dot\Psi_{\rm P} + \mu\dot\theta - \H,
\label{eq:lgc_tor_approx}
\end{align}

\noindent contains the time derivative of one variable too much. The standard approach in the present context is to take ${\bm P} = \{\P_\zeta,\P_\vartheta,\mu)$ and ${\bm \Theta} = \{\vartheta,\zeta,\theta\}$ to be the canonical action-angle coordinates and eliminate the term $(\rho_\parallel + \alpha)q\beta_*\dot\Psi$. Besides the methods outlined in the introduction Section~\ref{sec:intro}, it has been pointed out on p.~78 of Ref.~\onlinecite{WhiteTokBook3} that --- in the case of an axisymmetric field ($\alpha = \Phi = 0$) --- the component $\beta_*$ enters the equations for $\rho_\parallel$, $\Psi_{\rm P}$ and $\vartheta$ only via the denominator $D$ in Eq.~(\ref{eq:d}), so that omitting $\beta_*$ corresponds merely to a distortion of time. In the toroidal direction, the effect of $\beta_*$ also causes only a non-secular modulation of the GC speed that averages to zero during one poloidal transit. Thus, the unperturbed axisymmetric system in Boozer coordinates can be made canonical (so as to explicitly reveal its Hamiltonian character) by simply omitting $\beta_*$. Incidentally, once $\beta_*$ is ignored, even the perturbed GC Lagrangian attains the desired canonical form,
\begin{align}
\L_{\rm c} =\;& \P_\zeta \dot\zeta + \P_\vartheta \dot\vartheta + \mu\dot\theta - \H,
\label{eq:lgc_tor_d0}
\end{align}

\noindent with $\H(\rho_\parallel,\Psi_{\rm P},\vartheta,\zeta,t)$. The Euler-Lagrange equations ${\rm d}_t\partial_{\dot{\eta}}\L = \partial_\eta\L$ imply that Eq.~(\ref{eq:dpz_dt_expanded}) turns into Hamilton's equation
\begin{equation}
\dot\P_\zeta = -\left.\frac{\partial\H_{\rm c}}{\partial\zeta}\right|_{\P_\zeta,\P_\vartheta,\vartheta,t={\rm const}.}
\end{equation}

\noindent for a Hamiltonian in canonical form, $\H_{\rm c}(\P_\zeta,\P_\vartheta,\vartheta,\zeta,t)$,
\begin{equation}
\H_{\rm c} = (\P_\zeta + \Psi_{\rm P} - g\alpha)^2 \frac{B^2}{2g^2} - \frac{|{\bm v}_{\rm E}|^2}{2} + \mu B + \Phi,
\end{equation}

\noindent where the dependence on parallel velocity has been replaced by $\P_\zeta$ through $\rho_\parallel = (\P_\zeta + \Psi_{\rm P})/g - \alpha$, and all $\Psi_{\rm P}$-dependencies become functions of the two momenta via $\Psi_{\rm P}(\P_\zeta,\P_\vartheta)$. The full set of Hamilton's equations is
\begin{subequations}
\begin{align}
\dot{\P}_\vartheta = -\partial_\vartheta H_{\rm c}, \quad \dot{\P}_\zeta = -\partial_\zeta \H_{\rm c},
\\
\dot\vartheta = \partial_{\P_\vartheta}H_{\rm c}, \quad \dot\zeta = \partial_{\P_\zeta}\H_{\rm c},
\end{align}
\label{eq:he}
\end{subequations}

The present work was motivated by our preference to stay with the simpler and more intuitive formulation using {\it noncanonical} coordinates that are defined in terms of {\it unperturbed} fields only, like the set ${\bm \eta} = \{\rho_\parallel,\Psi_{\rm P},\vartheta,\zeta\}$ used in our code {\tt ORBIT}. What we have shown in this paper is that the retention of $\beta_*$ in the equations of motion for these coordinates preserves at least two properties of the Hamiltonian system --- (i) phase space conservation and (ii) energy conservation --- even in the presence of electromagnetic perturbations in the form of $\Phi$ and $\alpha$.

%\vspace{1cm}
%\bibliographystyle{unsrt}
%\bibliography{references}

\end{document}